\documentclass{aa}
\pdfoutput=1

\usepackage[usenames,dvipsnames]{color}
\usepackage{ulem}
\usepackage[usenames,dvipsnames]{color}
\usepackage{ulem}
\usepackage{graphicx}
\usepackage{txfonts}
\usepackage[font=footnotesize]{caption}
\usepackage[utf8]{inputenc}
\usepackage{subfigure}
\usepackage{chngcntr}
\usepackage{natbib}
\usepackage{rotating}
\usepackage{color}
\usepackage{IEEEtrantools}
\begin{document}
\newcommand{\blue}{\textcolor{blue}}
\newcommand{\red}{\textcolor{red}}

\title{An advanced scattered moonlight model for Cerro Paranal\thanks{Based on observations made 
             with ESO telescopes at Paranal Observatory}}

\author{A. Jones
          \inst{1}
\and
          S. Noll\inst{1}
\and
	  W. Kausch\inst{1}
\and
	  C. Szyszka\inst{1}
\and
	S. Kimeswenger\inst{2,1}
          }

   \institute{Institute for Astro and Particle Physics, Leopold Franzens Universit{\"{a}}t Innsbruck,
              Technikerstrasse 25, 6020 Innsbruck, Austria\\
		\email{Amy.Jones@uibk.ac.at}
   \and
   Instituto de Astronom{\'{i}}a, Universidad Cat{\'{o}}lica del Norte,
                Avenida Angamos 0610, Antofagasta, Chile
          }

   \date{Accepted for publication in A\&A October 2013}

\abstract{The largest natural source of light at night is the Moon, and it is the major contributor to the astronomical sky background. Being able to accurately predict the sky background, including scattered moonlight is important for scheduling astronomical observations.  We have developed an improved scattered moonlight model, in which the components are computed with a better physical understanding as opposed to the simple empirical fit in the frequently used photometric model of \citet{1991PASP..103.1033K}.  Our spectroscopic model can better trace the spectral trends of scattered moonlight for any position of the Moon and target observation.  This is the first scattered moonlight model that we know of which is this physical and versatile.  We have incorporated an observed solar spectrum, accurate lunar albedo fit, and elaborate scattering and absorption calculations that include scattering off of molecules and aerosols.  It was designed for Cerro Paranal, but can be modified for any location with known atmospheric properties.  Throughout the optical range, the uncertainty is less than 20\%.  This advanced scattered moonlight model can predict the amount of scattered moonlight for any given geometry of the Moon and target, and lunar phase for the entire optical spectrum.}


\maketitle

\section{Introduction}

The current trend in astronomy has been to build larger and larger telescopes.  The operating costs for running these large telescopes are high and careful planning of observations is needed for telescope time is always in demand.  This means that more accurate predictions and estimation of the sky background are needed to understand how long an exposure is necessary for a given observation with a certain signal to noise ratio.  The brightest natural source of light in the night sky, and therefore the greatest contributor to the sky background noise, is scattered moonlight.  Having a reliable model of the moonlight for sky background estimation is critical.  Also, many observers are trying to characterize faint objects spectroscopically, and knowing accurately the spectrum of the background would allow astronomers to predict which spectral features are observable within a given exposure time.  By improving the scattered moonlight model within a sky background model, we can increase telescope scheduling efficiency.

The long standing scattered moonlight model used by ESO (European Southern Observatory) was by \citet{1987Noao....10...16W} and provides a table of the magnitudes for five photometric bands of the night sky at five different moon phases.  It does not depend on the positions of the Moon or target observation and was measured during solar minimum.  This model is limited in producing a scattered moonlight spectrum which is accurate enough for the current and future telescope operations.
  
The current, widely used model was developed by \citet{1991PASP..103.1033K}, with 52 citations \citep[e.g.][]{2013NatSR...3E1722D,2013arXiv1307.6116K,2013MNRAS.432.3262T}.  It is an empirical fit to data in the V-band taken at the 2800 m level of Mauna Kea.  Even though the fit was separated into various specific functions, such as initial intensity from the Moon, Rayleigh and Mie scattering, it was still a parametrization based on only 33 observations in one photometric band specifically for Mauna Kea.  The accuracy is between 8 and 23\% if not near full Moon.  In a previous paper, \citet{2012A&A...543A..92N}, we presented a spectroscopic extension of the \citet{1991PASP..103.1033K} model.  It was optimized for Cerro Paranal and covered the optical regime.  The lunar albedo was taken to be constant with respect to wavelength and scaling factors for the different functions were introduced to better fit data from Cerro Paranal.

\begin{figure*}[th]
\centering
\includegraphics[width=0.95\textwidth]{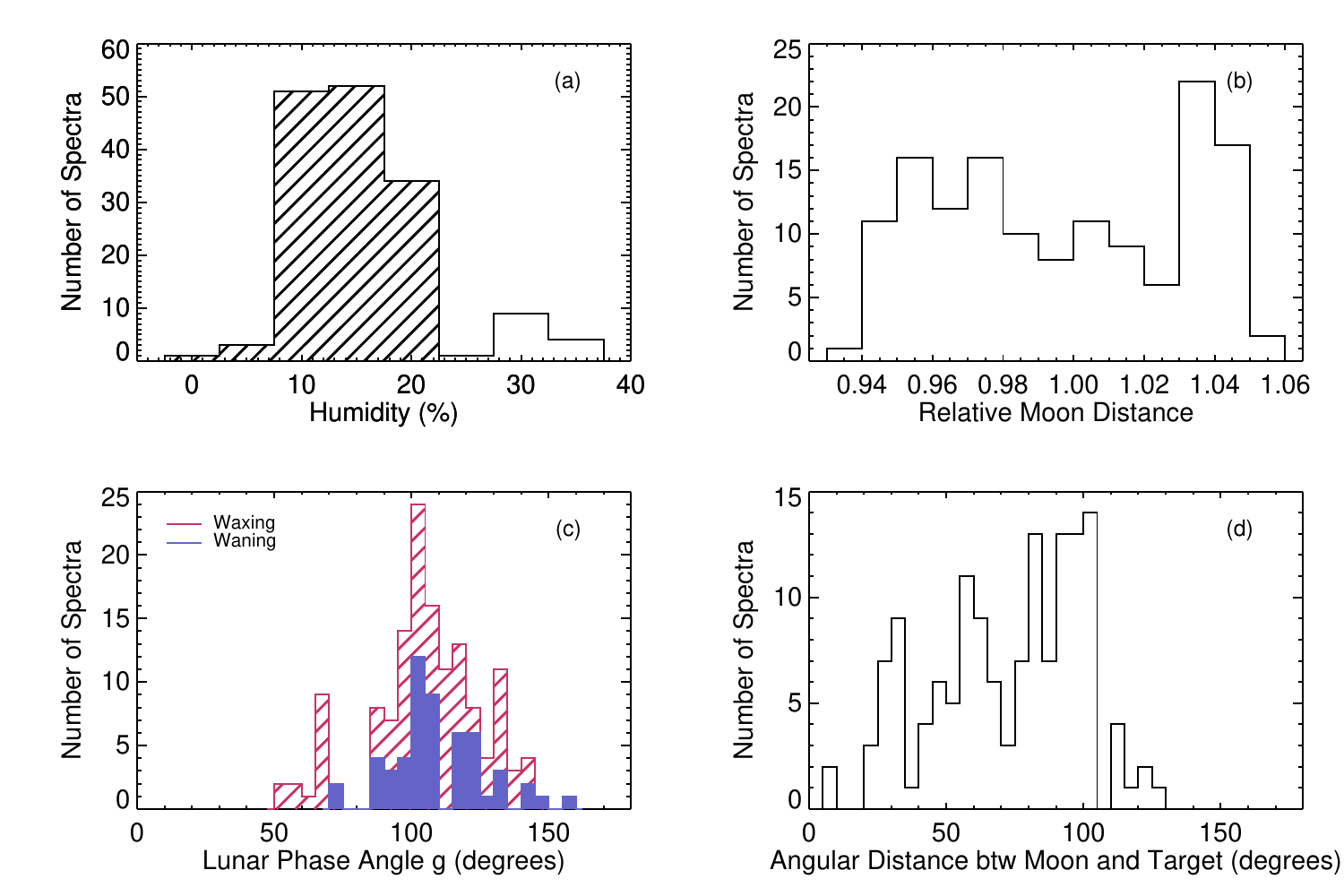}
\caption{Properties of the FORS1 data from \citet{2008A&A...481..575P}.  The data shown has been preselected to not have any clouds during observation and plots b, c, and d also have $\leq 20\%$ humidity. {\bf (a):}  The distribution of the amount of humidity at the time of observation as given by ESO Ambient Conditions Database.  Only observations with $\leq 20\%$ humidity were analyzed and is shown here as the filled in part of the histogram. {\bf (b):} Shows the distribution of the relative moon distance.  A moon distance of 1 corresponds to the average moon distance of 384,400 km. {\bf (c):} Here we have plotted the distribution of lunar phase angles $g$.  Over-plotted is the number of spectra taken during a waxing and waning Moon, which effects the solar selenographic longitude. {\bf (d):} The distribution of the angular distance between the Moon and target observation $\rho$ is displayed.  See Section 2.1 for more details.}
\label{plot_prop}
\end{figure*}

Our advanced scattered moonlight model works from the UV to the near-IR, but has only been fully tested in the optical, due to a current lack of data.  It has been calibrated and investigated with 141 optical spectra and has an overall accuracy of $\sigma \lesssim 0.2$ mag.  The model has been split into physically based modules which are given by either physical models or the best current fits.  The present version is optimized for Cerro Paranal, but can be modified for any location with information about its atmospheric properties.  Our model is fully 3D and can predict the amount of scattered moonlight for any configuration of the Moon and target.  It includes higher order scattering and is therefore still reliable at high zenith distances (when either object is near the horizon).  Because our scattered moonlight model produces a spectrum, it can be used for finding spectral features and trends as well as photometric magnitudes.

The original purpose of designing an advanced scattered moonlight model, as part of the full sky background model, was for the Austrian accession to ESO to improve the ETC (Exposure Time Calculator), and thus enhance telescope scheduling and efficiency.  Our optical sky background model, but with an older version of the scattered moonlight model, is described in \citet{2012A&A...543A..92N}.  The full sky background model estimates the amount of background light from 0.3 to 30 micron for Cerro Paranal.  It includes all relevant components, such as scattered moonlight and starlight, zodiacal light, airglow line emission and continuum, scattering and absorption within the Earth's atmosphere, and thermal emission from the atmosphere and telescope.  Each component was designed with the latest knowledge and results in the field and was thoroughly checked with archival ESO data.  The new scattered moonlight model is the topic of this paper.  The zodiacal light is found using the prescription from \citet{1998A&AS..127....1L} and the airglow model is based on local observations and semi-empirical modeling \citep{2008A&A...481..575P}.  The scattering and absorption are calculated with two codes, one that is fully 3D for the scattering and a line-by-line radiative transfer code for molecular absorption.  The scattering and absorption will also be described in this paper, in the context of the scattered moonlight.  The thermal emission is estimated as a gray body.  The new sky background model has already been implemented by ESO.

In this paper we will discuss the optical portion of the advanced scattered moonlight model.  In a subsequent paper, we will discuss the extension of this model into the near-IR using X-Shooter data \citep{2011A&A...536A.105V}.  For the remainder of the paper, the term "moon model" refers to the scattered moonlight model.  In Section 2, we describe the moon model.  In Section 3, we compare our scattered moonlight model with the observations and previous models, and finally in Section 4 we present our conclusions.



\section{The Model}

To accurately calculate the scattered moonlight, there are several components that must be considered.  The simplest way to account for the various pieces is to follow the path of light from the source to the instrument.  The source of the scattered moonlight is the Sun.  Then the light is reflected by the Moon which depends on the lunar albedo.  This is mainly a function of the lunar phase.  Next, the moonlight enters the Earth's atmosphere, where it is scattered by molecules (Rayleigh) and aerosols (Mie), and absorbed.  The light can be scattered multiple times, including off of the ground, before it reaches the telescope.  This depends on the properties of the atmosphere and the positions of the Moon and target observation.

In the following subsections, we will discuss each of these steps in more detail.  Section 2.1 introduces the calibration data set.  In Section 2.2 we discuss the solar spectrum, in Section 2.3 the lunar albedo, and in Section 2.4 a more general discussion of the set-up of the radiative transfer equations.  Then, we have a more detailed discussion about Rayleigh scattering in Section 2.5, Mie scattering in Section 2.6, and absorption in Section 2.7.

\subsection{Calibration Data Set}

We calibrated our full sky background model including the improved scattered moonlight model with a FORS1 (FOcal Reducer/low dispersion Spectrograph) data set from \citet{2008A&A...481..575P}.  FORS1 was a low resolution optical spectrograph at Cerro Paranal.  The data were taken between 1999 and 2005 with four different observing modes and consist of long-slit spectra, where plain sky could be extracted.  For the details about the different observing modes, see Table \ref{obs_mode}.  There are 1186 spectra, but only 26\% have moonlight.  For the analysis, we considered only 141 spectra with good weather conditions according to ESO's ambient weather conditions database\footnote{http://archive.eso.org/asm/ambient-server}.  The condition of good weather depended on two criteria, humidity and clouds.  First, we only considered spectra taken when the humidity was $\leq 20\%$.  Second, there must not be any detection of possible clouds in the night sky, within 5 hours of the observation.  Since we are measuring scattered moonlight, the entire sky must be free of clouds to ensure stable atmospheric conditions.  Fig. \ref{plot_prop} shows several properties of the data: the distribution of the humidity with the cutoff, moon distance, lunar phase angles with waxing and waning, and angular distance $\rho$.

\begin{table*}
\caption{Properties of the different observing modes in the FORS1 sky background data set\label{obs_mode}}
\centering
\begin{tabular}{cccccccc}
\hline\hline
\noalign{\smallskip}
Grism & Filter & Wav. Range & Resolution & Dispersion & Total \#  & \# with Moon & \# Used\\
 & & (\AA) & (\AA\;FWHM) & (\AA\;px$^{-1}$) & & & \\
\noalign{\smallskip}
\hline
\noalign{\smallskip}
300V & GG435 & 4300-8900\tablefootmark{a} & 12 & 2.6 & 676 & 188 & 70 \\
300V & $\ldots$ & 3615-8900\tablefootmark{b} & 12 & 2.6 & 163 & 36 & 24 \\
600B & OG590 & 3650-6050 & 5.3 & 1.2 & 143 & 29 & 18 \\
600R & GG435 & 5390-7530 & 4.5 & 1.0 & 207 & 60 & 29 \\
\noalign{\smallskip}
\hline
\end{tabular}
\tablefoot{Listed here is the total number of sky spectra, the number of sky spectra with moonlight, and the number of sky spectra with moonlight that was used in the analysis (Section 2.1).}
\tablefoottext{a}{order separation filter, fluxes $<4400$\AA\;were not used}
\tablefoottext{b}{second-order overlapping, fluxes $>6200$\AA\;were not used}
\end{table*}

\subsection{Solar Spectrum}

\begin{figure}[!ht]
\centering
\includegraphics[width=0.49\textwidth]{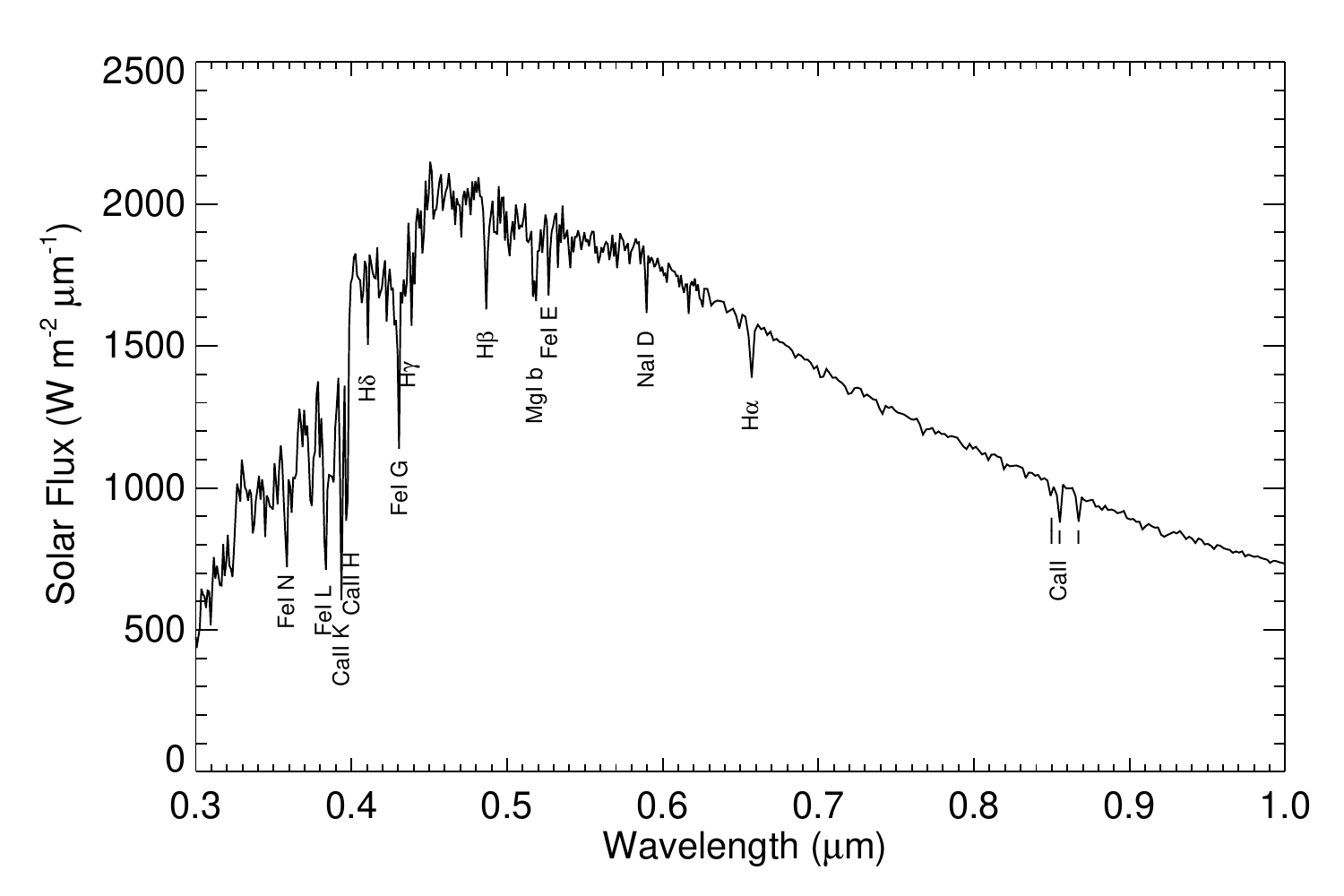}
\caption{Optical solar spectrum from \citet{1996AJ....112..307C} with principal absorption lines and bands labeled \citep{2011ApJS..195....6W}.  This was used as the source for our scattered moonlight model (Section 2.2).}
\label{sol}
\end{figure}

The source of the scattered moonlight is the Sun.  We use the spectrum from \citet{1996AJ....112..307C}, which covers wavelengths from 0.12 to 2.5 micron.  The UV and optical portion are from measurements from satellites and ground based data.  The near-IR spectrum is from data by the NASA CV-990 aircraft with a model spectrum.  The optical and near-IR spectrum has an uncertainty of $\leq 5\%$, while the UV has an uncertainty of $\sim 20\%$ due to solar variability, with a resolution of 1 to 2 nm.  Fig. \ref{sol} shows the optical solar spectrum.  It resembles the black body spectrum with typical absorption lines for a G2 star. 

\subsection{Lunar Albedo}

The lunar albedo ($A$) determines the amount of sunlight that is reflected off the lunar surface towards the Earth.  It is directly related to the intensity of light that will enter the Earth's atmosphere, called $I^*$.  They are related by \citep{2005AJ....129.2887K},
\begin{equation}
I^*=I_\mathrm{sol}\frac{\Omega_M}{\pi}A\Big(\frac{384,400}{M_\mathrm{dis}}\Big)^2.
\label{istar}
\end{equation}
$I_\mathrm{sol}$ is the spectrum of the light intensity of the Sun (see Section 2.2), and $\Omega_M$ is the solid angle of the Moon, where we take $\Omega_M=6.4177\times 10^{-5}$ sr.  For our set of sky observations \citep{2008A&A...481..575P}, we found the moon distances ($M_\mathrm{dis}$) with the online tool called JPL Horizons\footnote{http://ssd.jpl.nasa.gov/?horizons}.  The distance of $384,400$ km is the average moon distance.

The lunar albedo $A$ depends on several factors, including lunar phase, solar selenographic longitude, and wavelength.  We use an empirical fit based on the ROLO survey \citep{2005AJ....129.2887K}.  They used over 100,000 images of the Moon in 32 different photometric bands in the optical and near infrared for certain lunar phases.  The fit given by \citet{2005AJ....129.2887K} is,
\begin{IEEEeqnarray}{rCl}
\ln{A_\lambda}=\sum^3_{i=0} & a_{i,\lambda} & g^i+\sum^3_{j=1}b_{j,\lambda}\Phi^{2j-1}+d_{1,\lambda}e^{-g/p_1}\nonumber\\ && + d_{2,\lambda}e^{-g/p_2}+d_{3,\lambda}\cos{[(g-p_3)/p_4]}.
\label{lnalb}
\end{IEEEeqnarray}
The lunar phase is described by the parameter $g$, where $g=0^\circ$ corresponds to a full Moon and $g=180^\circ$ is a new Moon.  $\Phi$ is the solar selenographic longitude which describes whether the Moon is waxing or waning.  Because the maria are not uniformly distributed on the lunar surface, the overall reflectivity of the Moon varies depending on which portion of the Moon is illuminated.  We used the Astronomical Ephemeris by NASA\footnote{http://eclipse.gsfc.nasa.gov/TYPE/ephemeris.html} to find whether the Moon was waxing or waning.  In Eq. \ref{lnalb} the parameters $a_{i,\lambda}$, $b_{j,\lambda}$, and $d_{x,\lambda}$ for the 32 different wavelengths $\lambda$ are provided in Table 4 and the constants $p_n$ are in Eq. 11 of \citet{2005AJ....129.2887K}.
  
The last three terms in Eq. \ref{lnalb} are to correct for the opposition effect (exponential functions) and to minimize the residuals between the data and fit (cosine function).  The opposition effect is where the Moon has an enhanced brightness near full Moon \citep{1969NASSP.201...38W}.  We have neglected the terms that depend on the observer selenographic latitude and longitude.  They have a small impact, maximum of $\sim 7\%$ over a full Saros (libration) cycle, and are variables that are not easily available in astronomical data.

\begin{figure}[!ht]
\centering
\includegraphics[width=0.49\textwidth]{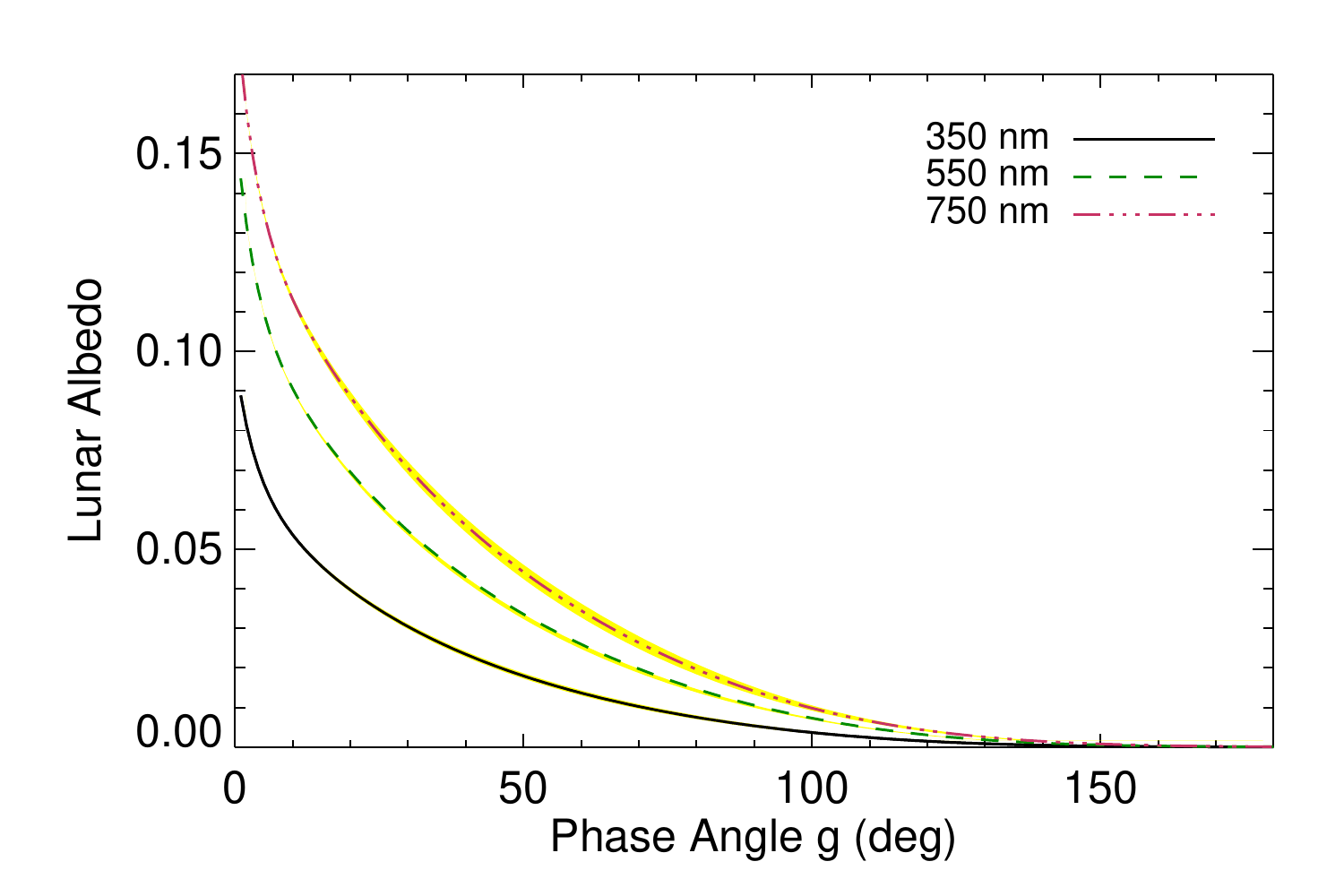}
\caption{Lunar albedo based on the \citet{2005AJ....129.2887K} fit as a function of the phase angle $g$, with the extrapolation where $g>97^\circ$, for three different wavelengths.  The variation due to the solar selenographic longitude is shown in yellow.  See Section 2.3 for more details.}
\label{plot_lnalb_g}
\end{figure}

The ROLO fit only covers the phase angles between $g=1.55^\circ$ and $97^\circ$.  We extrapolated this fit to $g=180^\circ$.  For the terms that depend on $\Phi$, when $g > 97^\circ$ we took the values of $\Phi$ at $g=97^\circ$.  This physically means that we are assuming that the ratio of maria to highlands is the same when $g>97^\circ$ as $g=97^\circ$.  The other terms in Eq. \ref{lnalb} are still well behaved when $g > 97^\circ$ and fit reasonably well with our data \citep{2008A&A...481..575P}.  It is expected from previous studies that the general dependence of $A$ on $g$ is exponential for all values of $g$ (outside the opposition effect) \citep[e.g.][]{1973AJ.....78..267L}, which we see in our extrapolation as well.  As $g$ increases the overall flux from the Moon decreases, and becomes less important compared with the other components of the sky background.  Thus, we believe the extrapolation to higher $g$ is reasonable.  In Fig. \ref{plot_lnalb_g}, we show the dependence of the lunar albedo on moon phase including the extrapolation when $g > 97^\circ$.  Also shown is the effect of $\Phi$ which is a bit larger at redder wavelengths and is over-plotted as a band of values in yellow.  At $g=97^\circ$, $\Phi$ can cause deviations of $\sim 6.8\%$.

\begin{figure}[!ht]
\centering
\includegraphics[width=0.49\textwidth]{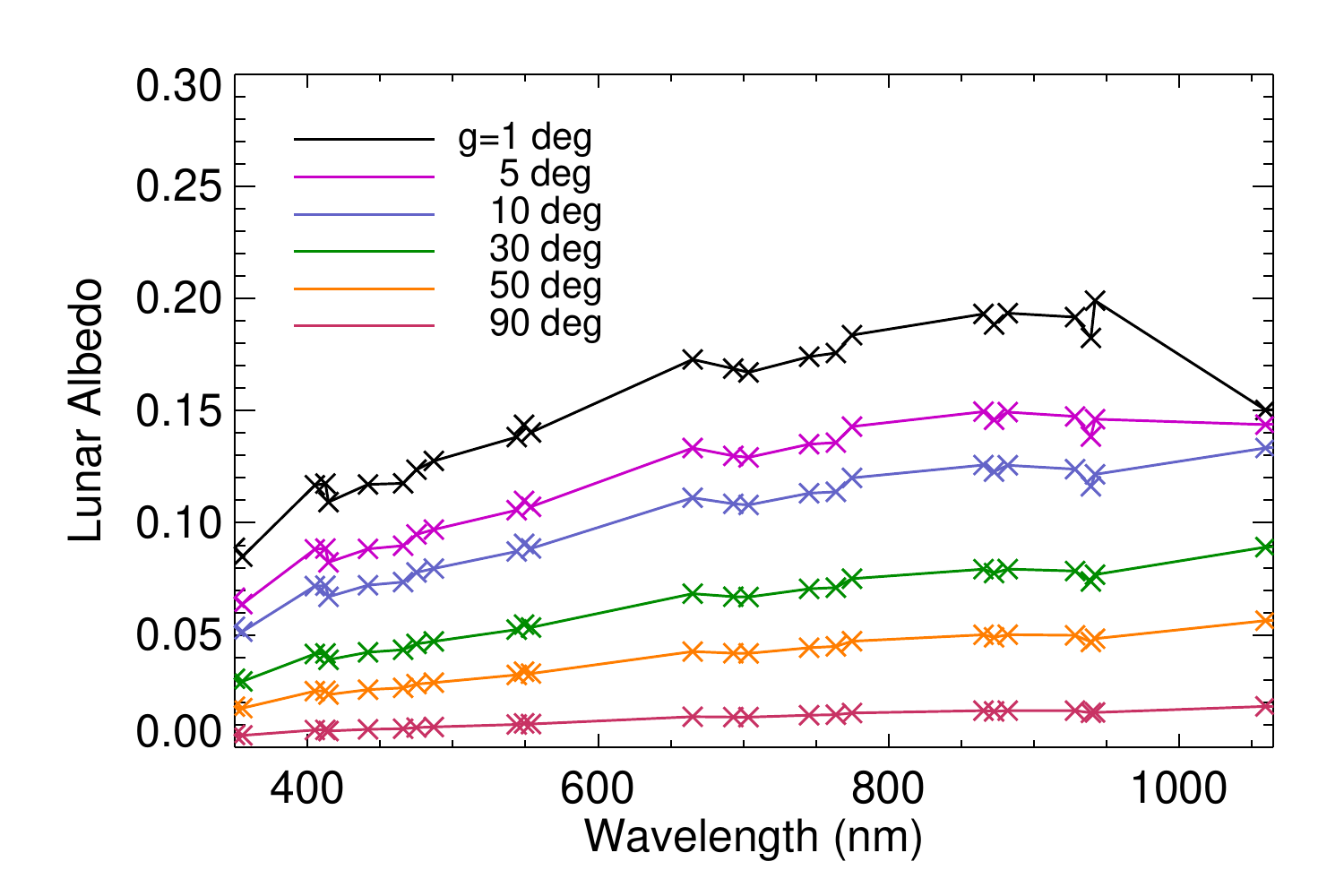}
\caption{Values of the lunar albedo given by ROLO \citep{2005AJ....129.2887K} presented as tick marks and the interpolation for the lunar albedo as a function of wavelength for several lunar phases $g$ (Section 2.3). }
\label{plot_lnalb_wav}
\end{figure}

The fit from the ROLO data was only done photometrically, however we require a full spectrum of $I^*$ for the scattered moonlight model.  To attain a fit for the entire optical spectrum, we performed a simple linear interpolation between the fit values provided.  This interpolation is shown in Fig. \ref{plot_lnalb_wav} for several different moon phases.  In general the wavelength dependence is fairly smooth and becomes less significant towards larger $g$.  There are some small variations where the photometric values are clustered.

According to \citet{2011Icar..214...30V}, there are some discrepancies in the overall flux calibration amongst the various Moon observations.  They find that the ROLO data at 603 nm is $13\%$ too faint.  They used an interpolation of the ROLO fit to find the albedo at 603 nm to compare it with their observations.   Since there is no information about the wavelength dependence for this correction, we divide the albedo at all wavelengths by $0.87$. 

We currently neglect the effects of polarization.  We assume that the change in the resulting scattered moonlight due to polarization is smaller than our current errors.  In general the polarization of the lunar surface is small, less than $10\%$ at maximum.  However, the maria can have a decent amount of polarization ($\sim 30\%$) \citep{1971A&A....10...29D}.


\subsection{Radiative Transfer}

\begin{figure}[!ht]
\centering
\includegraphics[width=0.49\textwidth]{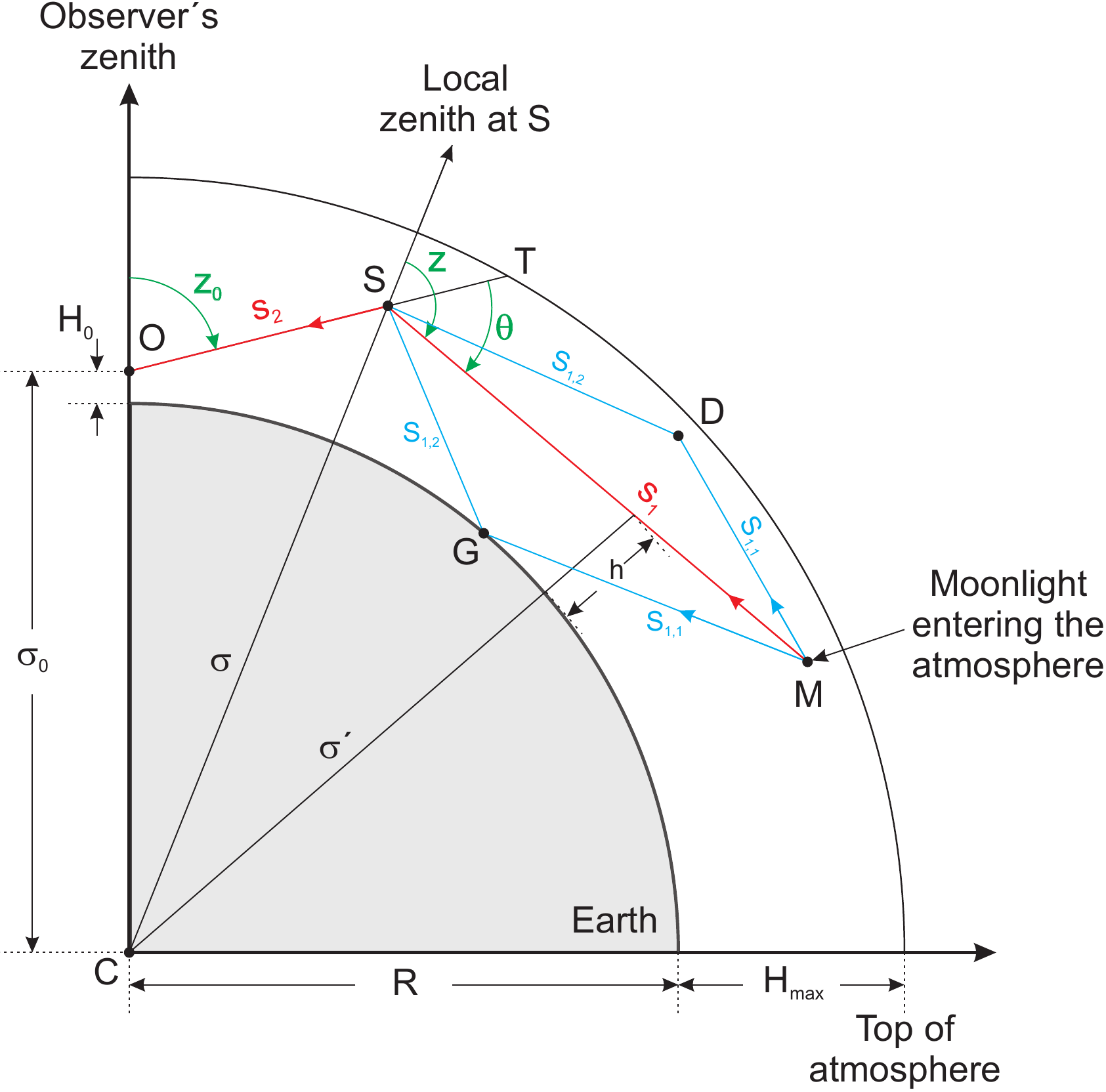}
\caption{Geometry of the scattering in the Earth’s atmosphere \citep[cf.][]{1967ApJ...147..255W,2012A&A...543A..92N}.  Point $M$, where the moonlight enters the top of the atmosphere, is not in the same plane as the other points. The azimuth of $M$ as seen from $S$ is $A_M$ (not shown).  $S$ and $T$ are at an azimuth $A_0$ (also not shown) for an observer at $O$.  Single scattering is shown at point $S$ along paths $s_1$ and $s_2$.  Double scattering occurs at point $D$ or off the ground $G$ (Section 2.4).}
\label{plot_scatsketch}
\end{figure}

The light from the Moon can be scattered many times or absorbed in the Earth's atmosphere before reaching the telescope.  We have developed a fully 3D single scattering code with estimates for double and higher order scattering.  At the various scattering points, an effective airmass is calculated for absorption.

The transmission of light $t$ can be directly related to the optical depth at zenith $\tau_0$ or the extinction coefficient $k$ and airmass $X$ by the following, 
\begin{equation}
t(\lambda)=e^{-\tau_0(\lambda)X}=10^{-0.4k(\lambda)X}.
\label{ext}
\end{equation}
We will use this formalism in the following sections to describe the effects of scattering.

\subsubsection{Single Scattering}

The single scattering can be done fully in 3D \citep{1967ApJ...147..255W,1975A&A....39..325S,2002ApJ...571...56B}.  We wish to obtain the integrated scattered light towards the azimuth $A_0$ and zenith distance $z_0$.  We take the Moon to be a point source.  We consider the scattering path elements $S$ of density $n(\sigma)$, with $\sigma$ being the radius vector from the center of Earth $C$ to $S$.  $S$ is located between the top of atmosphere $T$ to the observer $O$, at height $H_0$ above the surface (see Fig. \ref{plot_scatsketch}). The distance between $O$ and $C$ is $\sigma_0 = H_0 + R$, where $R$ is the radius of Earth (6371 km for the mean radius). For each path element $S$ at distance $s_2$ from $O$, the contributions of radiation from $M$, where the moonlight enters the atmosphere, to the intensity at $S$ along $s_1$ are considered.  This intensity includes the scattering of light out of the path (and possible absorption), which depends on the effective column density of the scattering/absorbing particles.  This is given by,

\begin{equation}\label{eq:effcoldens}
B_1(z, \,\sigma) = \int_{0}^{s_1(z, \,\sigma)} n(\sigma') \,\mathrm{d}\sigma'.
\end{equation}

The scattering intensity also depends on the wavelength-dependent extinction cross section $C_\mathrm{ext} (\lambda)$ of the various processes, Rayleigh, Mie and absorption. They are related to the optical depth $\tau$ (see Eq. \ref{ext}) and Eq. \ref{eq:effcoldens} at the zenith by,

\begin{equation}\label{eq:crosssection}
C_\mathrm{ext}(\lambda) = \frac{\tau_0(\lambda)}{B_0(0, \,\sigma_0)}.
\end{equation}

After scattering at $S$ the light travels along path $s_2$ and is extincted by $B_2(z_0,\sigma)$.  The total scattering intensity per solid angle $I_\mathrm{scat}$ at $(A_0,z_0)$ for the flux of the Moon $F^*$, related to $I^*$ (Eq. \ref{istar}), at $(A_{M},z_{M})$ as seen by $O$, for Rayleigh and Mie scattering separately, is given by,


\begin{eqnarray}\label{eq:scatintegral}
I_\mathrm{scat}(A_0, z_0) = \frac{C_\mathrm{scat}(\lambda)}{4\pi}\int_0^{s_2(z_0,\sigma_0)}n(\sigma)
\,P(\theta)\nonumber\\ \times F^*(A_{M},z_{M})\exp^{-\tau}\mathrm{d}s,
\end{eqnarray}
where
\begin{eqnarray}
\label{cext}
\tau=(C_\mathrm{ext,R}+C_\mathrm{ext,A})\times[B_\mathrm{1,R}(z,\sigma)+B_\mathrm{2,R}(z_0,\sigma_0)] \nonumber\\
+C_\mathrm{ext,M}\times[B_\mathrm{1,M}(z,\sigma)+B_\mathrm{2,M}(z_0,\sigma_0)].
\end{eqnarray}
Eq. \ref{cext} is for Rayleigh (R), Mie (M), and absorption (A).
In Eq. \ref{eq:scatintegral}, $C_\mathrm{scat}$ is the wavelength-dependent scattering cross section, which will deviate from $C_\mathrm{ext}$ if absorption occurs. The scattering phase function $P$ depends on the scattering angle $\theta$, the angle between the paths $s_1$ and $s_2$, and is related to the zenith distances and azimuths at $S$ by,
\begin{equation}
\label{anglgeo}
\cos{\theta}=\cos{\tilde{z}_0}\cos{\tilde{z}_{M}}+\sin{\tilde{z}_0}\sin{\tilde{z}_{M}}\cos{(\tilde{A}_0-\tilde{A}_{M})}.
\end{equation}
Here, $\tilde{z}_0$, $\tilde{A}_0$, $\tilde{z}_M$, and $\tilde{A}_M$ are the zenith distances and azimuths of the target and Moon at $S$.

In Eq. \ref{cext}, the various $C_\mathrm{ext}$ are calculated by Eq. \ref{eq:effcoldens} for their respective densities.  $C_\mathrm{ext,A}$ is calculated assuming the same particle distribution as Rayleigh scattering.  This is optimal for molecular oxygen absorption.

We neglect the effect of polarization on the scattering phase function.  Scattered moonlight does have some degree of polarization, similar to scattered sunlight, and has large areas of the sky with very little polarization \citep{1998NW.....85..333H,2001JGR...10622647G}.  The distribution of polarized light is not fully understood and would be difficult to implement.  

For the vertical distribution of the scattering molecules, we use the standard barometric formula,
\begin{equation}\label{eq:baroform}
n(h) = n_0\exp{(-h/h_0)}.
\end{equation}
Here, $h = \sigma - R$, the sea level density $n_0 = 2.67 \times 10^{19}$ cm$^{-3}$, and the scale height $h_0 = 7.99$ km above the Earth’s surface \citep{1975A&A....39..325S,2002ApJ...571...56B}.  For the troposphere and the lower stratosphere, where most of the scattering occurs, this is a good approximation. Cerro Paranal is at an altitude of $H_0 = 2.64$ km.  For the thickness of the atmosphere, we take $H_\mathrm{max} = 200$ km.  For Rayleigh scattering, $C_\mathrm{scat} = C_\mathrm{ext}$, i.e. no absorption is involved.  

For the height distribution of aerosols, we also use Eq. \ref{eq:baroform} with $n_0 = 1.11 \times 10^4$ cm$^{-3}$ and $h_0 = 1.2$ km. This is the tropospherical distribution of \citet{1966ApOpt...5.1769E}.  Dust, in particular soot, also absorbs radiation, thus $C_\mathrm{scat}$ is lower than $C_\mathrm{ext}$.  We used the OMI (Ozone Monitoring Instrument) satellite data to find the median ratio of 0.97 (range between 0.90 and 0.99) for the area around Cerro Paranal \citep{2006ITGRS..44.1093L}.  The model is not very sensitive to changes in the molecular or aerosol distribution, because we scale them with the extinction curve.

\subsubsection{Double Scattering}

The above scattering equations only deal with single scattering.  We also considered double scattering.  For this, we divided the path $s_1$ into two different possible paths $s_{1,1}$ and $s_{1,2}$ where scattering and absorption can occur at the path elements $D$ (for double scattering), in between the two paths (see Fig. \ref{plot_scatsketch}).  At $D$ the scattering is treated in a similar way as in Eq \ref{eq:scatintegral}, but over the additional path elements.  To simplify this integral and speed up the computing time, we made an approximation that allowed the integration to become an analytical expression with percent level accuracy (Eq. \ref{eq:effcoldens}).  We assumed that the Moon coordinates where absolute coordinates with respect to the observer, such that the path $s_2<<R$ and the Moon is sufficiently far enough away to disregard slight angular shifts to the Moon coordinates for the different scattering points $S$ and $D$.  Near the horizon and when scattering occurs at a far distance from the telescope, this assumption breaks down.  However, most of the scattering occurs in the lower troposphere and a few km from the observer (with mean distances of $\sim10$ km).  The calculation was performed along a grid, which exponentially grows with distance.  Where $\theta$ is small and therefore the contribution from the forward scattering peak of aerosols is high, we increased the resolution of the grid.  For the higher order scattering, including double scattering, absorption by molecules via an effective airmass was no longer considered.

\begin{figure}[!ht]
\centering
\includegraphics[width=0.49\textwidth]{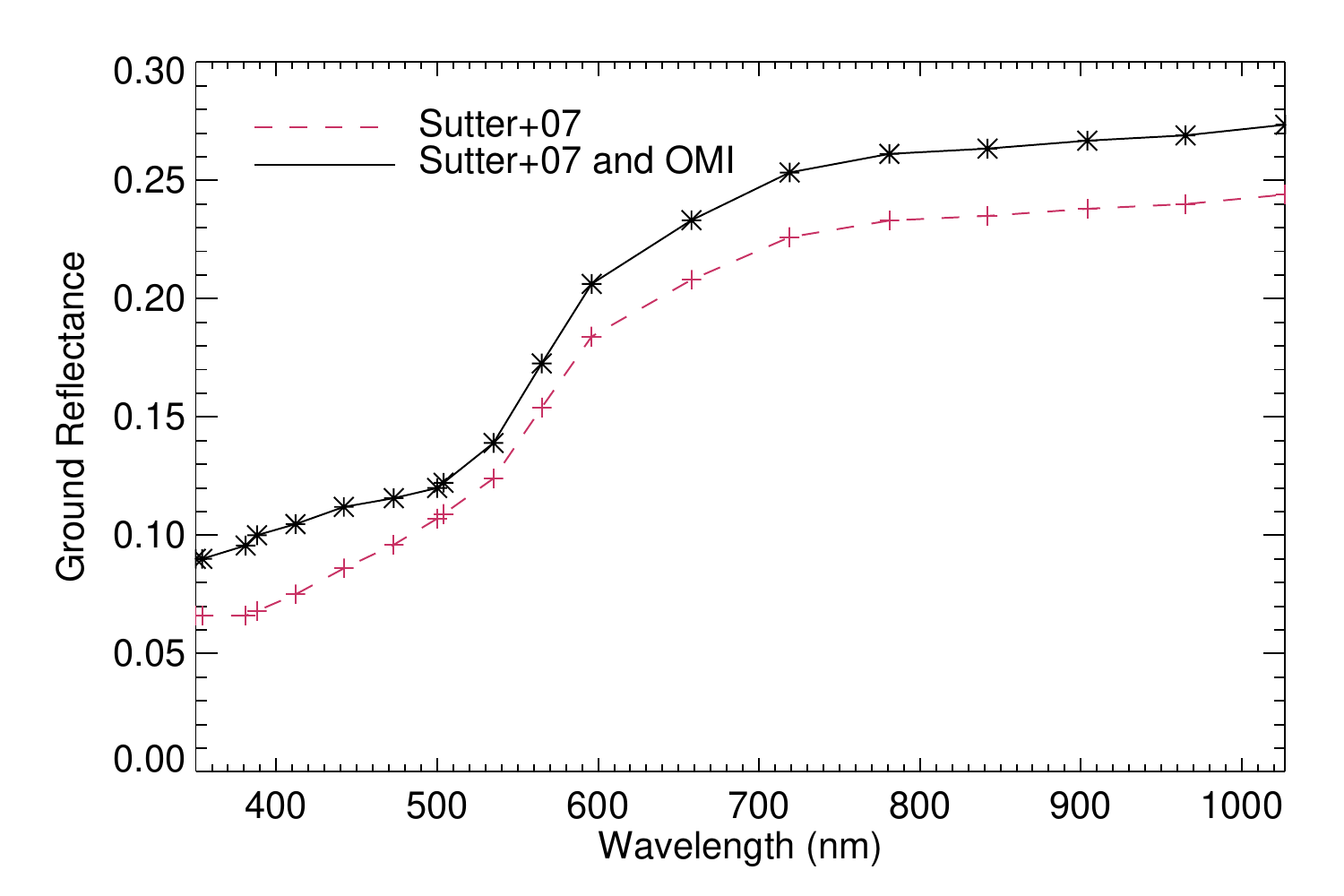}
\caption{Wavelength-dependent ground reflectance used for double scattering (see Section 2.4 for more details).}
\label{plot_gr}
\end{figure}

In the case of double scattering, scattering can also occur at the ground, labeled as point $G$ in Fig \ref{plot_scatsketch}.  Since Cerro Paranal is located at $2\;635$ m in a mountainous region, we took $G$ to be at a height of 2 km, although the exact height has little influence.  To estimate the ground reflection, we used the A1 profile from \citet{2007JGRG..112.4S10S} based on soil samples taken in the Chilean desert.  We then scaled it to the values from OMI \citep{2006ITGRS..44.1093L} at the three provided wavelengths for the region around Cerro Paranal.  We followed the procedure of \citet{2008JGRD..11318308K} to obtain a proper average ground reflection, since there are many errors associated with this calculation.  This gave us a wavelength dependent ground reflection at our desired location, as shown in Fig. \ref{plot_gr}.

Double scattering is the sum of two different single scattering events, which can be off of molecules, aerosols, or the ground.  We sum the contributions from the three different types of scattering to arrive at a final double scattering intensity, called $I_{DS}$.  Because the moon model was intended for the ETC, it needed to be computationally quick and the exact position of the Moon is irrelevant.  For this purpose, we made additional approximations for $I_{DS}$.  We computed weighted averages over both the zenith distance to the target $z_0$ and to the Moon $z_M$.  The weighting for $z_0$ was based on all the sky spectra from \citet{2008A&A...481..575P} and for $z_M$ from theoretical modeling of the positions of the Moon.  $I_{DS}$ is then only a function of the angular separation between the Moon and target $\rho$ and the optical depth $\tau$.  The average uncertainty for this simplification is on the order of 5\%, and is higher at large zenith distances but quickly decreases towards smaller zenith distances.

\subsubsection{Multiple Scattering}

We now consider the intensity for multiple scattering, anything higher than two, called $I_{MS}$.  The intensity for single scattering will be labeled as $I_{SS}$.

For $I_{MS}$, we compared the contributions for $I_{DS}$ with $I_{SS}$.  We assumed that the ratio between $I_{DS}$ and $I_{SS}$ would be the same for each consecutive order of scattering.  We then summed over the geometric series and multiplied it by $I_{DS}$,
\begin{equation}
I_{MS}=I_{DS}\Big(\frac{1}{1-I_{DS}/I_{SS}}-1\Big).
\end{equation}
We imposed an upper limit for $I_{DS}/I_{SS}$ of 0.9, so $I_{MS}$ would not diverge.  Similar to the weighted averages for $I_{DS}$, we calculated a weighted average over the difference between the azimuths for the target and Moon $A_0-A_M$, which were weighted evenly.  With Eq. \ref{anglgeo}, this gave a weighted average over the angular separation $\rho$.  In the end, $I_{MS}$ is only a function of optical depth $\tau$, which can be related to wavelength.

The total scattering intensity for the Moon $I_\mathrm{tot}$ is simply the sum of the contributions from $I_{SS}$, $I_{DS}$, and $I_{MS}$.  With the weighting over the various parameters, we created a table of correction factors $f$ for the added amount of scattering due to $I_{DS}(\tau,\rho)$ and $I_{MS}(\tau)$.
\begin{equation}
I_\mathrm{tot}(\tau,\rho,z_0,z_M)=I_{SS}(\tau,\rho,z_0,z_M)f(\tau,\rho).
\end{equation}

\begin{figure}[!ht]
\centering
\includegraphics[width=0.49\textwidth]{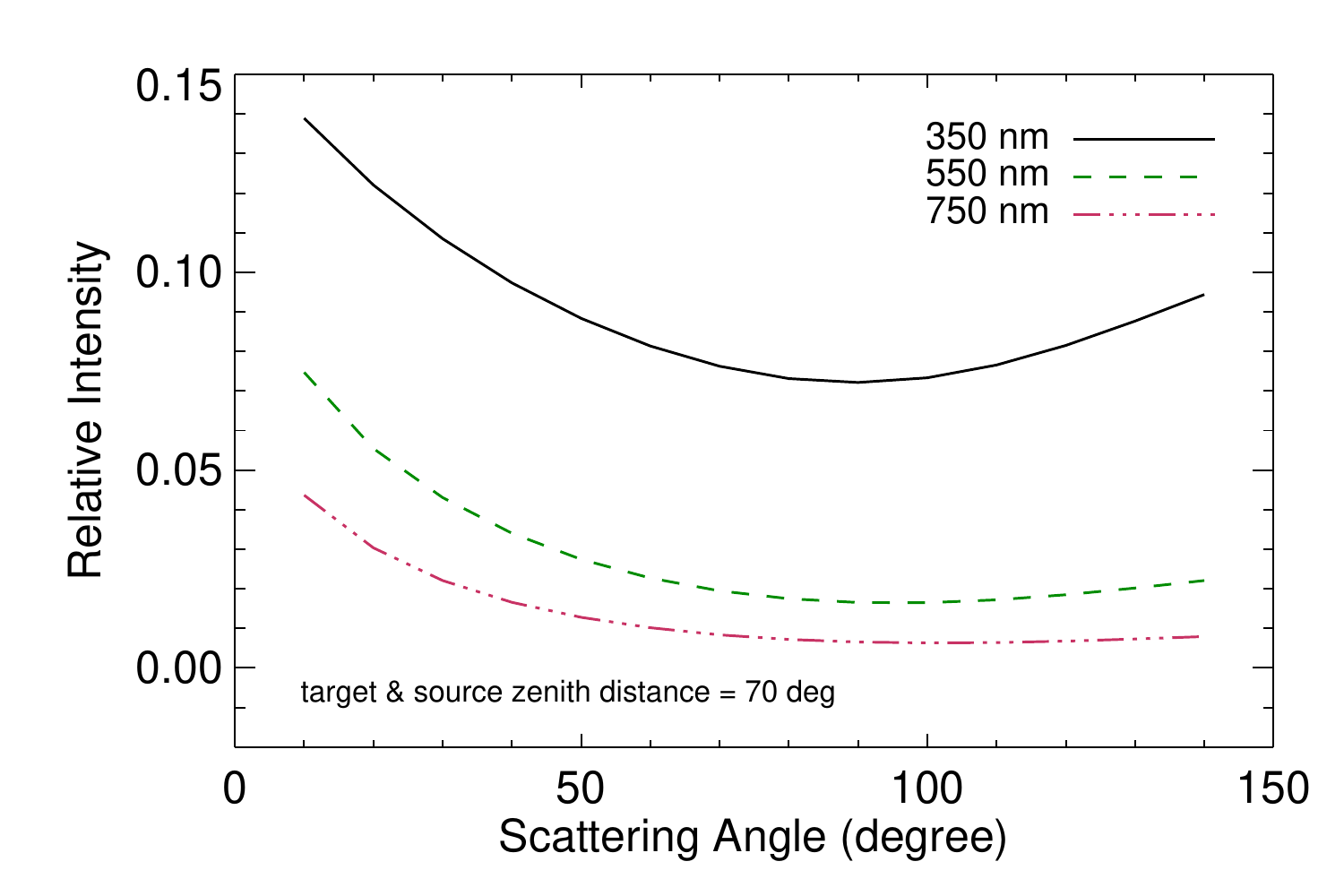}
\caption{Relative intensity to the light entering the atmosphere for scattering as a function of the scattering angle for a given geometry at three different wavelengths.  The bluer wavelength is dominated by Rayleigh scattering and the redder wavelength by Mie scattering (Section 2.4).}
\label{plot_scat}
\end{figure}

The higher order scattering, $I_{DS}$ and $I_{MS}$, typically contribute only a few percent and is strongest at high optical depth.  Some typical scattering curves are shown in Fig. \ref{plot_scat} for three different wavelengths.  At the redder wavelengths, Mie scattering is dominating and the scattering function approaches an exponential.  At bluer wavelengths, the characteristic Rayleigh scattering phase function is the most influential.  

\begin{figure}[!ht]
\centering
\includegraphics[width=0.49\textwidth]{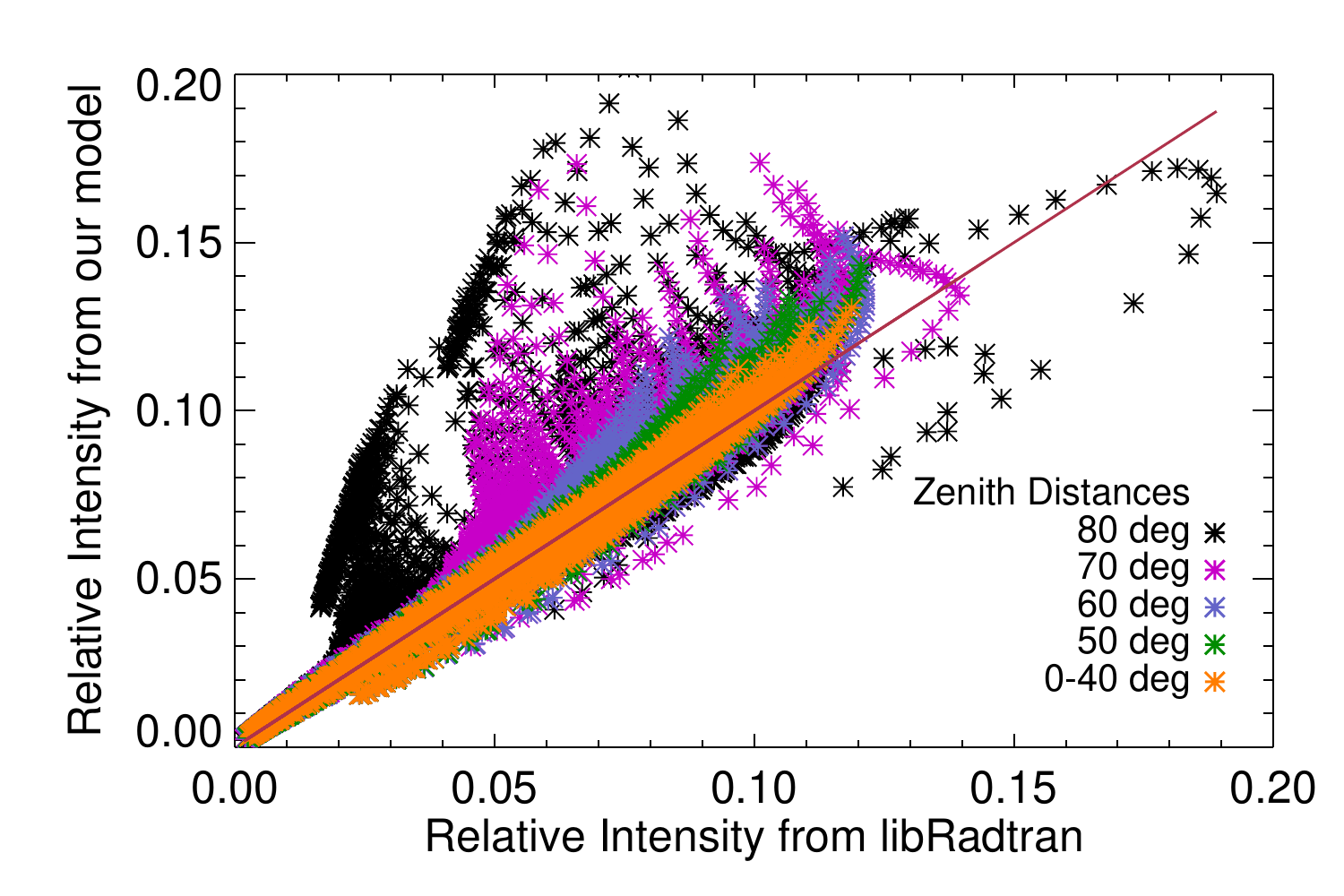}
\caption{Comparison of our scattering code used in the advanced scattered moonlight model with libRadtran (commonly used radiative transfer code), including single, double and multiple scattering.  In different colors we have plotted the relative intensities for each based on the maximum zenith distance of the Moon (source) or target (observer).  When both zenith distances are below $50^\circ$ the agreement is excellent.   See Section 2.4 for more details.}
\label{plot_libeffcomp}
\end{figure}

We have compared our scattering intensities $I_\mathrm{tot}$ with a radiative transfer code called libRadtran \citep{2005ACP.....5.1855M} for a grid of zenith distances of the Moon and target, and the angular separation $\rho$ in $10^\circ$ steps for many different optical depths (corresponding to the full optical wavelength range).  We have made this comparison to check the validity and accuracy of our scattering code.  libRadtran is a widely used third party code for calculating radiative transfer in the atmosphere.  It has several different numerical solvers, including the default which uses the plane parallel approximation.  With this approximation, libRadtran is not accurate when either the source (Moon) or target are near the horizon.  For all of the cases, 85\% of $I_\mathrm{tot}$ from our code and libRadtran agreed within a relative error of 20\%.  For $I_\mathrm{tot}$ with zenith angles less than $70^\circ$, 75\% agreed to within 10\% relative error.  The comparison between our code and libRadtran is shown in Fig. \ref{plot_libeffcomp}.  With zenith angles $\leq 50^\circ$, there is very good agreement.  As expected, at higher zenith angles where libRadtran is no longer accurate, the two codes diverge.

In the next three subsections, we will describe how Rayleigh, Mie and absorption are treated within the context of the scattering and absorption equations.



\subsection{Rayleigh Scattering}

The scattering off of molecules in the Earth's atmosphere can be well described by Rayleigh scattering, which assumes the particles are much smaller than the wavelength.  Rayleigh scattering is characterized with having a steep dependence on wavelength $\sim \lambda^{-4}$ and at Cerro Paranal it is quite stable \citep{2012A&A...543A..92N}.  For the extinction due to Rayleigh scattering we use the following parametrization from \cite{2002aiar.book.....L}, 
\begin{equation}
\tau_R(\lambda)=\frac{P}{1013}(8.6\times 10^{-3}+6.5\times 10^{-6}H)\;\lambda^{-(3.9+0.074\lambda+0.050/\lambda)}.
\label{ray}
\end{equation}
Here, $\lambda$ is in $\mu$m, and $P$ is the pressure at a height $H$ of the observer, which we take to be 744$\pm 1.5$ hPa and 2.64 km, respectively \citep{2012A&A...543A..92N}.

The phase function for Rayleigh scattering is also well defined and is given by,
\begin{equation}
P(\theta)=\frac{3}{4}(1+\cos{\theta}^2),
\label{rayph}
\end{equation}
where $\theta$ is the scattering angle.

The Rayleigh scattering is taken at each path element $S$ and $D$, and is a well described and stable component of the model.

\subsection{Mie Scattering}

For the aerosol scattering, we have gathered information from several different sources and tried to build the most physical description.  We take the \AA ngstr\"{o}m Law fit derived by \citet{2011A&A...527A..91P} for Cerro Paranal in the optical wavelengths.  We then decomposed this fit into several aerosol types given by \citet{2012acc.book.....W}, and produced the phase function using a Mie scattering code for log normal distributions based on \citet{1983asls.book.....B}.

A typical way to describe the optical depths $\tau$ and wavelength dependence $\lambda$ of aerosols in the optical is with the \AA ngstr\"{o}m Law, $\tau=\beta\lambda^{-\alpha}$, where $\alpha$ and $\beta$ are the fitting parameters \citep{1929GA.....11..156A}.
The fit by \citet{2011A&A...527A..91P} is based on 600 observations of 8 different spectrophotometric standard stars taken with FORS1.  The data were taken within a short time span of 6 months in 2009, so there could be some unknown, long-time variations.  They compared the leftover aerosol signature with what was predicted by the radiative transfer code LBLRTM (Line-By-Line Radiative Transfer Model) \citep{2005JQSRT..91..233C}.  The tropospheric aerosols (height $<10$ km) needed to be scaled down by 25\% in LBLRTM to match the observations.  The optical depth at Cerro Paranal is very low and the aerosols are almost purely background aerosol, with no urban component \citep{2011A&A...527A..91P}.  In the optical, their fit is given by,
\begin{equation}
k_{aer}=k_0\lambda^a,
\end{equation}
where $k_0=0.013\pm0.002$ and $a=-1.38\pm 0.06$ mag airmass$^{-1}$, valid from 0.4 to 0.8 $\mu$m.  We use the best fit $k_0$ value quoted in \citet{2011A&A...527A..91P} of 0.014 mag airmass$^{-1}$.  The aerosol component can fluctuate significantly, as can be seen in data taken at Mauna Kea \citep{2013A&A...549A...8B}.  They have the \AA ngstr\"{o}m exponent of $a=-1.3\pm 1.4$ and state that at 330 nm it can vary as much as 0.4 mag airmass$^{-1}$ during photometric nights.

\begin{figure}[!ht]
\centering
\subfigure[]{\includegraphics[width=0.49\textwidth]{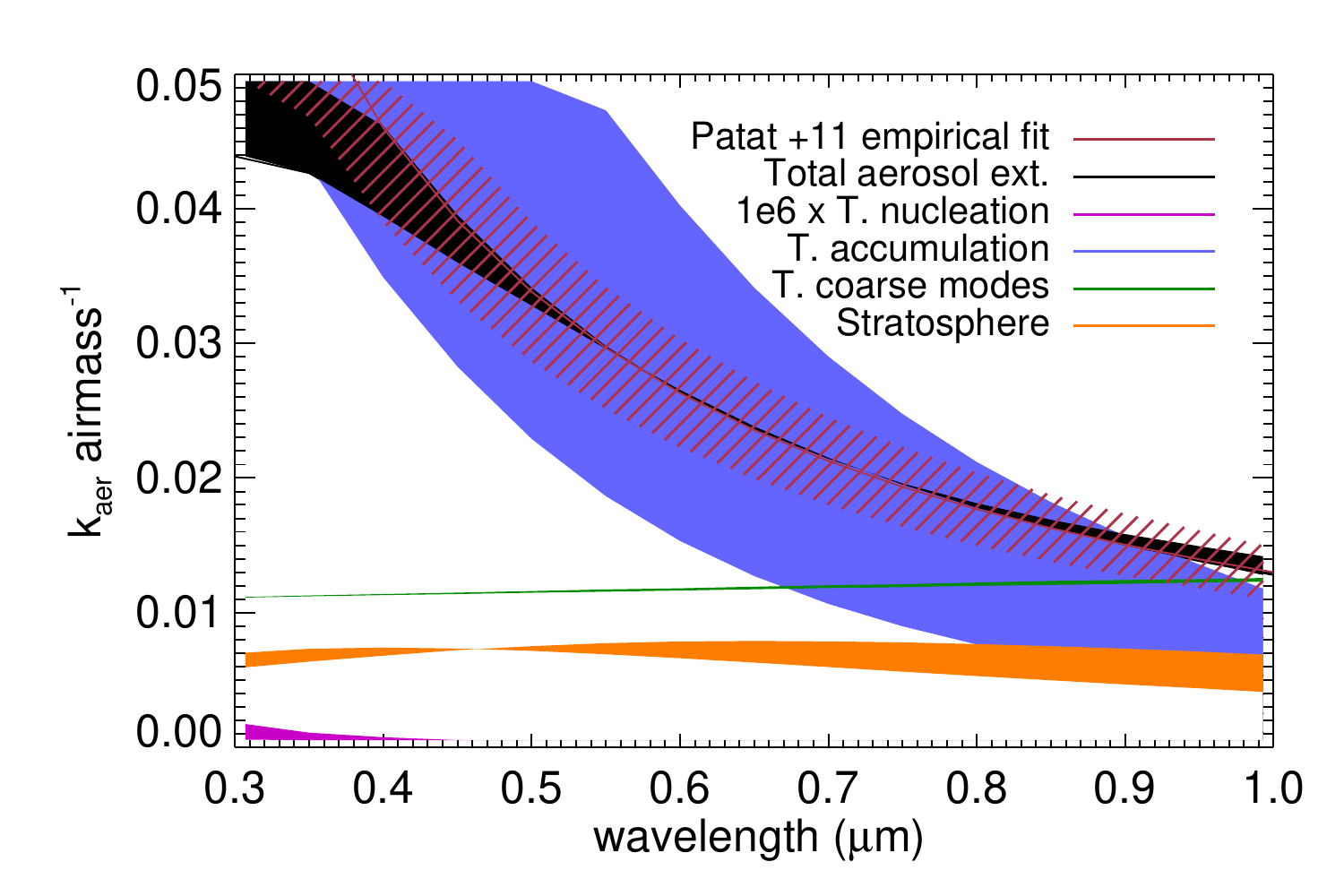}}\hfill
\subfigure[]{\includegraphics[width=0.49\textwidth]{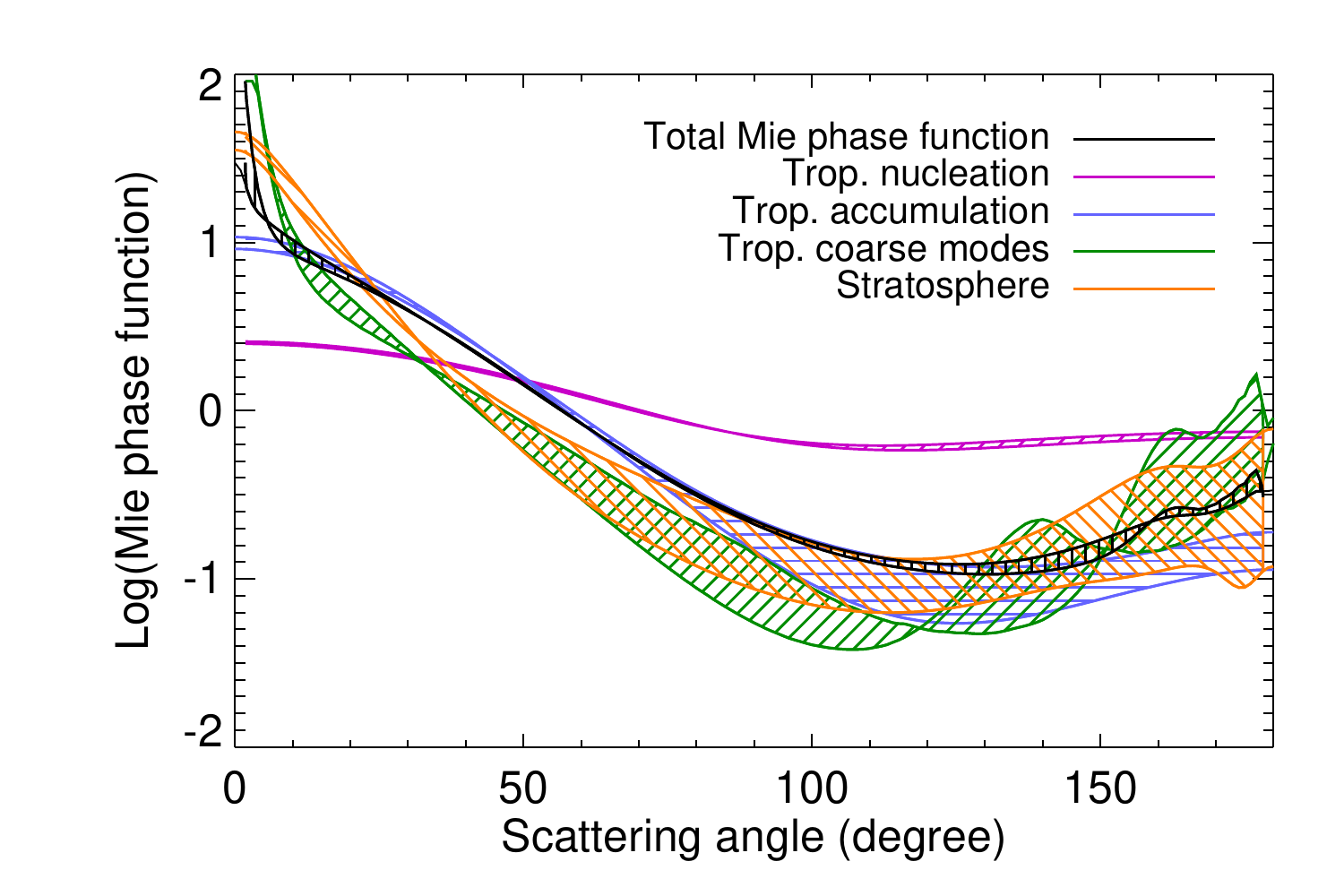}}\\
\caption{Different properties of the various types of aerosols. In both plots the width of the line is due to the uncertainty of the refractive index ($N$ ranges between 1.3 and 1.5) {\bf (a):}  Shows the extinction coefficient of Mie Scattering at 500 nm.  The range for the total Mie extinction curve are given by the following values for tropospheric nucleation, accumulation, and coarse, and stratospheric modes, respectively: 100, 100, 60, and 60\% for $N=1.3$ and 100, 45, 5, and 100\% for $N=1.5$.  The latter is what we have used for the analysis.  Also plotted is the \citet{2011A&A...527A..91P} fit with the reported errors.  The total Mie extinction curve (after scaling the various components) is in good agreement with the fit in the optical regime. {\bf (b):} This shows the Mie phase functions for the different modes and the total Mie phase function.  See Section 2.6 for more details.}
\label{plot_mie}
\end{figure}

We are interested in extending the scattered moonlight model into the infrared, as well as wanting a more physical basis for the aerosol extinction to more accurately calculate the phase function.  Therefore, we decomposed the \AA ngstr\"{o}m Law.  We used the size distributions given in \citet{2012acc.book.....W} for the remote continental tropospheric and stratospheric aerosols. The tropospheric aerosols are split into three types: nucleation, accumulation, and coarse modes.  Nucleation consists of newly produced particles from either direct emissions of combustion products or gas-phase condensation reactions.  Accumulation is from coagulation of nucleation particles, condensation of products from the gas-phase chemical reactions onto particles, and chemical reactions in the aqueous phase of clouds.  The mass of the coarse mode comes from mineral dust, sea salt, and biogenic material.  The stratospheric aerosols are in a layer around 25 km above sea level and consist mostly of sulfuric acid particles with an admixture of nitrosyl sulfates and solid granules containing silicates \citep{2012acc.book.....W}.  We use a log normal distribution given by,
\begin{equation}
\frac{dN(r)}{d\log{r}}=\frac{1}{\sqrt{2\pi}}\frac{n}{\log{s}}\exp{\Big[-\frac{(\log{r/R})^2}{2(\log{s})^2}\Big]}.
\label{lognormal}
\end{equation}
$N$ is the cumulative number density distribution in particle cm$^{-3}$, and $r$ is the particle radius in $\mu$m. Table \ref{tab_mie} provides the values for the parameters $n$, $R$, and $\log{s}$ of the various aerosol modes, which represent the number of particles, average radius, and a distribution parameter, respectively.  To calculate the extinction curve for each aerosol type we used a Mie scattering IDL code for log normal distributions written by G. Thomas\footnote{http://www.atm.ox.ac.uk/code/mie/mie\_lognormal.html} based on \citet{1983asls.book.....B} for single particles \citep{2004ApOpt..43.5386G}.  With the extinction curves for the various aerosol types along with the fit for Cerro Paranal, we determined how much each curve needed to be scaled to match the overall fit.  However, there are degeneracies.  The refractive index $N$ is not known for these background aerosols, so we varied it between 1.3 and 1.5. Varying $N$ changes the scaling of the different aerosol components needed to match the fit.  There is also a degeneracy between two of the aerosol types, remote continental tropospheric coarse mode and stratospheric, because their optical extinction curves are similar.  In Fig. \ref{plot_mie} (a) we have plotted the various Mie extinction curves.  The extinction from the different aerosol types with a range in $N$ between 1.3 and 1.5 are shown, along with the total curve that is consistent with the observed one.  It can easily be seen that the tropospheric nucleation mode has little to no effect, whereas the tropospheric accumulation has the greatest influence.

With the current FORS1 data set \citep{2008A&A...481..575P}, there is little to no difference in the overall results between $N=1.3$ or $1.5$ and the various amounts of the different aerosol components.  We chose $N=1.5$, with 45\% tropospheric accumulation, 5\% tropospheric coarse modes, and 100\% stratospheric aerosols, where the total matches the \citet{2011A&A...527A..91P} fit.  The tropospheric nucleation contributed extremely little to the overall aerosol scattering, and so was left at 100\%.  Since Cerro Paranal is located at $2\,635$ m, we lowered the amount of tropospheric aerosols and kept all the stratospheric particles.

With the log normal Mie scattering code we also calculated the Mie phase function of the different aerosol components and the overall Mie phase function.  This method of determining the phase function gives a realistic forward scattering peak, but can poorly approximate the back scatter.  This is caused by the fact that the aerosol particles are not spherically symmetric, and therefore the scattering phase function tends to flatten out at higher angles, rather than dip to a minimum as in the Mie approximation \citep[e.g.][]{2006AS..37..1287H}. Since the forward scattering dominantes the back scatter by a few orders of magnitude, we prefer the Mie scattering phase function over the Henyey-Greenstein \citep{1941ApJ....93...70H} approximation, which depends on the average asymmetry parameter of the aerosols.  Additionally, the majority of the stratospheric aerosols in this region are sulfates, which are in droplets, and so are fairly round.  The amount of coarse mode aerosol is low, which is the least round type of aerosol.  This will also provide another method to break the degeneracies amongst the various aerosol modes.  Finally, in astronomical observations it is rare to have the target and Moon at angular distances $>120^\circ$, where the back scatter peak becomes important.  In Fig. \ref{plot_mie} (b), the phase functions for the different aerosol types are plotted along with the total Mie phase functions.  Once again each mode is shown with a range of $N$ between 1.3 and 1.5.


\begin{table}
\caption{Aerosol modes \label{tab_mie}}
\centering
\begin{tabular}{cccc}
\hline\hline
\noalign{\smallskip}
Type & n & R & log s\\
 & cm$^{-3}$ & $10^{-1}$ $\mu$m & $10^{-1}$ \\
\noalign{\smallskip}
\hline
\noalign{\smallskip}
Trop nucleation & 3.20 $\times 10^3$ & 0.10 & 1.61  \\
Trop accumulation & 2.90 $\times 10^3$ & 0.58 & 2.17 \\
Trop coarse modes & 3.00 $\times 10^{-1}$ & 9.00 & 3.80 \\
Stratospheric & 4.49 $\times 10^0$ & 2.17 & 2.48 \\
\noalign{\smallskip}
\hline
\end{tabular}
\tablefoot{The values used for Mie scattering of remote continental aerosols from \citet{2012acc.book.....W} (Section 2.6).}
\end{table}

\subsection{Absorption}

The absorption is calculated using the radiative transfer code LBLRTM \citep{2005JQSRT..91..233C} with a merged atmospheric profile for Cerro Paranal and the effective airmass $X_\mathrm{eff}$ at point $O$ calculated from the single scattering at $S$.  LBLRTM is a widely used radiative transfer code in atmospheric sciences and it uses the atomic line database HITRAN (High-Resolution TRANsmission) \citep{2009JQSRT.110..533R}.  This gives the amount of absorption from the various atmospheric molecules.  It requires an atmospheric profile, and we use a merged one, which combines the temperature, pressure, and chemical composition as a function of height from three different sources.  We combined MIPAS (Michaelson Interferometer for Passive Atmospheric Sounding) equatorial profile \citep[see][]{2010A&A...524A..11S} with one from GDAS (Global Data Assimilation System) \footnote{http://ready.arl.noaa.gov/gdas1.php}, and finally scale the quantities of the lower atmosphere to the local meteorological data taken at Cerro Paranal.  The MIPAS profile contains height information on 30 molecular species.  The GDAS database, which is based on measurements, provides modeled profiles of temperature, pressure, and the relative humidity up to a height of about 26 km on a 3 hour basis with a spatial grid of $1^\circ\times 1^\circ$.  The local meteorological data from the ESO MeteoMonitor \footnote{http://archive.eso.org/asm/ambient-server} gives the local temperature, relative humidity, and pressure.  We currently use bimonthly mean, merged profiles for all the nights.

For each single scattering point $S$, an effective airmass $X_\mathrm{eff}$ relative to the direct path from the zenith to observer is calculated to scale the amount of absorption due to the given geometry of the Moon and target.  We do not calculate $X_\mathrm{eff}$ from the higher order scattering, since this contribution is quite small.  $X_\mathrm{eff}$ then modifies the transmission from the moonlight as calculated by LBLRTM.

For determining $X_\mathrm{eff}$, we compare the intensities with and without extinction, $I^{(+)}$ and $I^{(-)}$, respectively.  We first look at a specific path element for a given $S$.  We find the column density of that path element for both Rayleigh $B_{n,\mathrm{R}}$ and Mie scattering $B_{n,\mathrm{M}}$ (similar to Eq. \ref{eq:effcoldens}).  Then with $C_\mathrm{scat}$ and phase function $P(\theta)$, we calculate $I^{(-)}$ by,
\begin{equation}
I^{(-)}=\sum^n I^{(-)}_n,
\end{equation}
where,
\begin{equation}
I^{(-)}_n=C_\mathrm{scat,R} B_{n,\mathrm{R}} P_\mathrm{R}(\theta_n)+C_\mathrm{scat,M} B_{n,\mathrm{M}} P_\mathrm{M}(\theta_n).
\end{equation}

To calculate $I^{(+)}$, we use $I^{(-)}$ and $\tau$ from Eq. \ref{cext}, and is given by,

\begin{equation}
I^{(+)}=\sum^n I^{(-)}_n \exp{(-\tau_n)}.
\end{equation}
Here, $\tau_n$ is given at a reference wavelength, where the absorption is strong and the scattering is weak.  Within this constraint, we arbitrarily choose a wavelength of 1 $\mu$m for the scattering at the zenith, which gives $\tau_\mathrm{R+M}=0.02$.

Then with $I^{(+)}$ and $I^{(-)}$ along with the optical depth at zenith $\tau_0$, $X_\mathrm{eff}$ is,
\begin{equation}
X_\mathrm{eff}=\frac{\log{(I^{(-)}/I^{(+)})}}{\tau_0}.
\end{equation}
Here, $\tau_0$ is the reference $\tau_\mathrm{R+M}=0.02$ plus $\tau_\mathrm{A}$.

Finally, the scaled scattering intensity $I_\mathrm{scat,A}$ for absorption is given by,

\begin{equation}
I_\mathrm{scat,A}=I_\mathrm{scat}\exp{(-\tau_\mathrm{A} X_\mathrm{eff})}.
\end{equation}

The three main absorbing molecules in the optical are O$_2$, H$_2$O, and O$_3$.  LBLRTM \citep{2005JQSRT..91..233C} with the merged atmospheric profile produces a transmission curve for the absorption.  We use the particle distribution from Rayleigh scattering as an approximation for the absorbing molecules, which is optimal for O$_2$.  The water vapor, however, is distributed differently from the main air molecules.  The absorption from H$_2$O mostly occurs at low altitudes.  Since the distribution of particles is scaled with the extinction curve, the amount of H$_2$O is fairly accurate.  Changing the profile did not significantly change the resulting modeled spectrum.

The ozone is treated separately using the van Rhijn formula with a height of 25 km, which is given by,
\begin{equation}
X_\mathrm{oz}=\frac{1}{\sqrt{1-0.992\sin^2{z_M}}}.
\label{rhijin}
\end{equation}
Here $X_\mathrm{oz}$ is the effective airmass for ozone and $z_M$ is the zenith distance to the Moon.  There are two assumptions made.  The first one is that we treat the ozone only at a height of 25 km and not at a distribution of heights.  The thickness can vary with season and geographic location.  This assumption allows for a much simpler calculation of $X_\mathrm{oz}$.  The second is that we are only considering the light that is coming along the direct path from the Moon, by using the Moon's zenith distance.  We assume that the majority of the light has not been scattered prior to passing through the ozone layer.  Since most of the scattering takes place in the lower troposphere, which is distinctly below 25 km, this appears to be a safe assumption.  Following \citet{2011A&A...527A..91P}, we multiply the amount of ozone given in the standard atmospheric profile by 1.08.

\section{Comparison with FORS1 Data and Previous Models}

We have compared our model with the FORS1 data (see Section 2.1 for details) \citep{2008A&A...481..575P} and with the previous moon model from \citet{2012A&A...543A..92N} based on \citet{1991PASP..103.1033K}.  We found that overall, for the complete sky background model, an uncertainty of $\sigma \sim0.2$ mag with the FORS1 data set.  For the moon model, the uncertainty is $\sim 0.15$ mag.

\begin{figure}[!ht]
\centering
\subfigure[]{\includegraphics[width=0.49\textwidth]{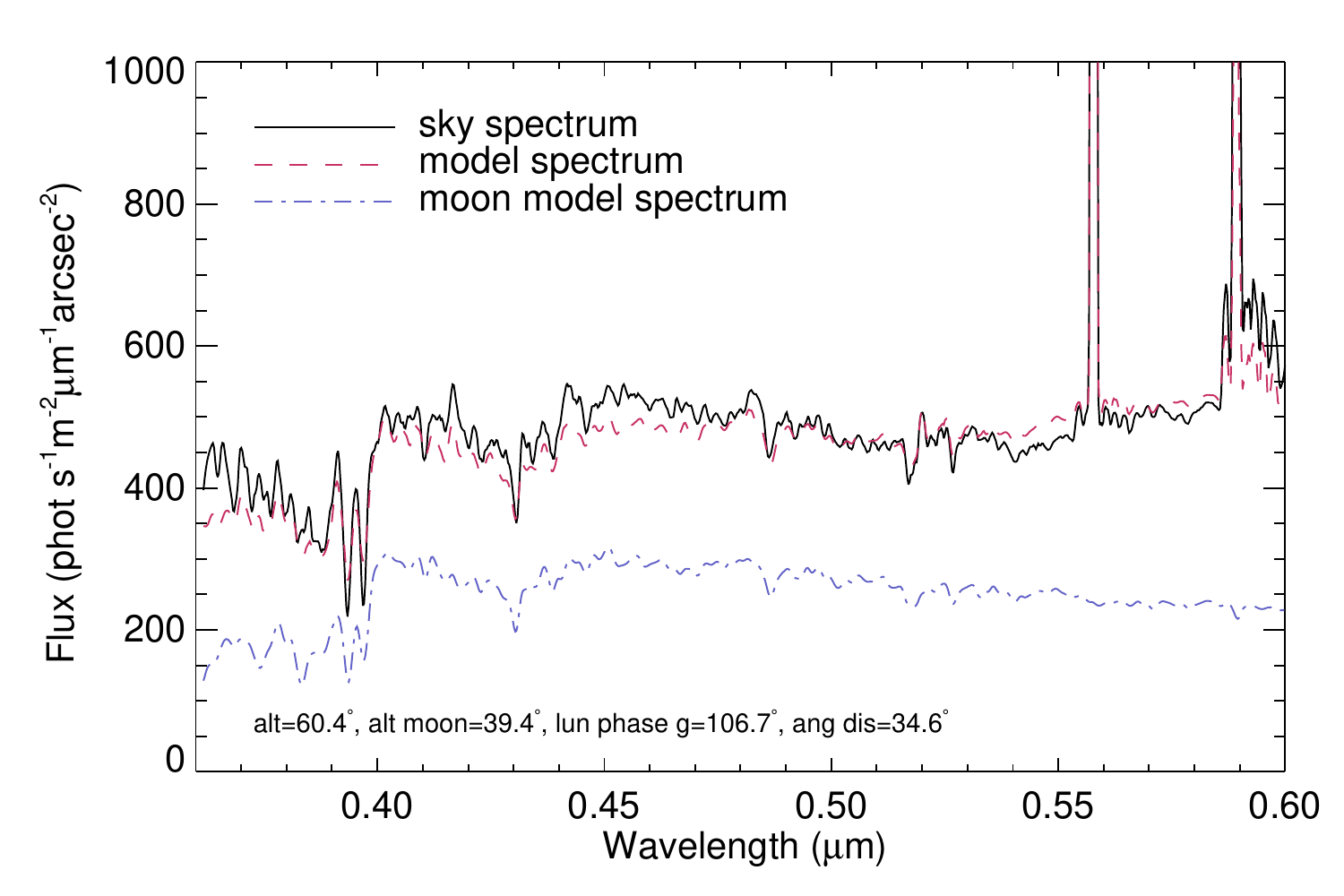}}\hfill
\subfigure[]{\includegraphics[width=0.49\textwidth]{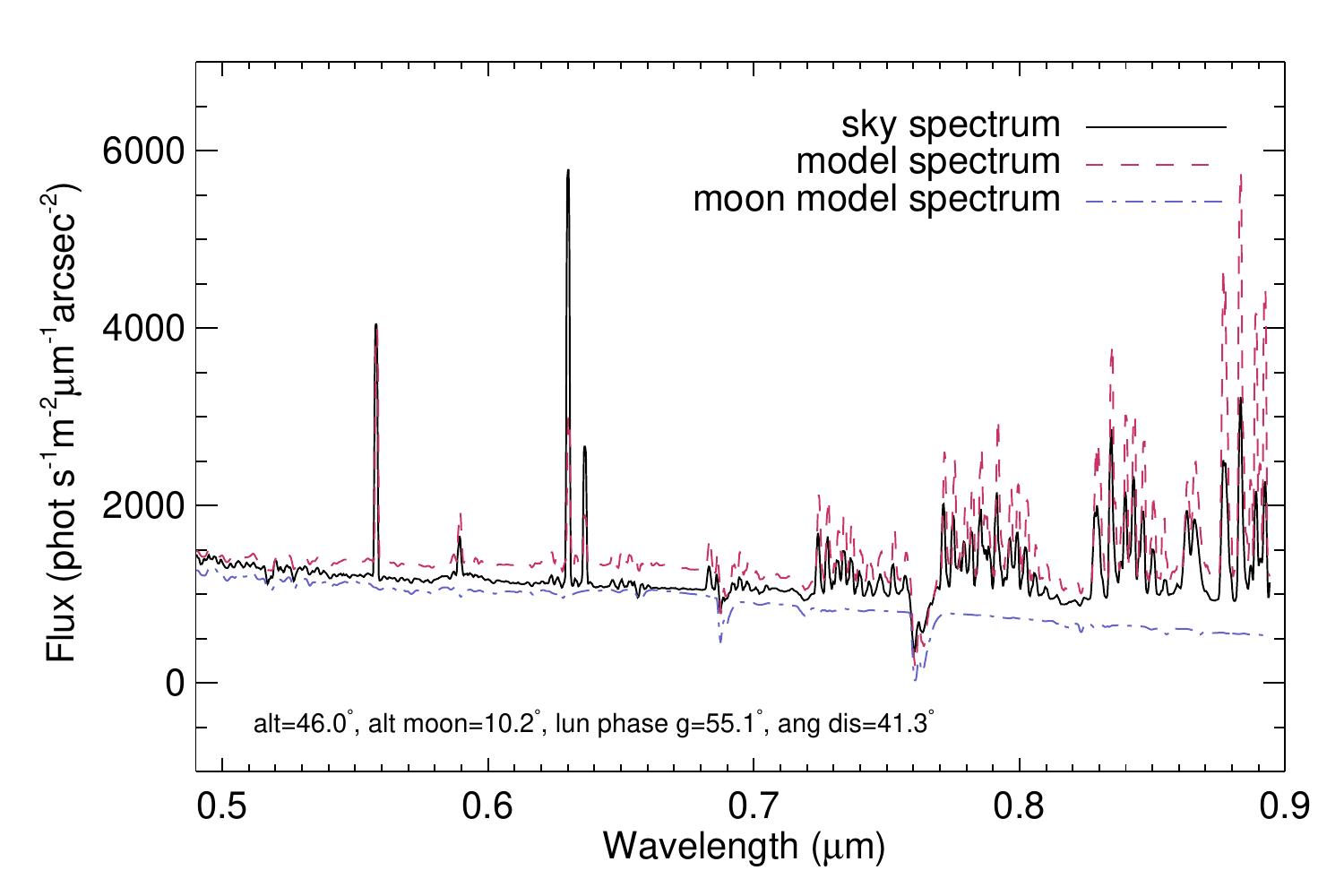}}\\
\caption{Two examples of the quality of the model.  The observed sky spectrum is shown in black with the full sky background model in red.  The scattered moonlight portion of the model is given in blue.  In both cases the model fits the observed spectrum fairly well. {\bf (a):}  This observation was done with moderate amount of moonlight.  The moonlight comprises roughly half of the overall sky background flux. {\bf (b):} This one was with a significant amount of moonlight, as can be seen by the increase in overall flux.  The scattered moonlight completely dominates the amount of background light.  See Section 3 for more details.}
\label{plot_fors1}
\end{figure}

Fig. \ref{plot_fors1} shows two examples of how well the model fits the observed data.  The spectrum in Fig. \ref{plot_fors1} (a) has moderate moonlight and is around 1st quarter and (b) has significant moonlight and is near full Moon.  The full sky model is consistent with the observed spectrum in both scenarios.  Also shown is the scattered moonlight model which is a large portion of the overall sky background flux.  For (b), the model tends to slightly overestimate the amount of background light.

\begin{figure}[!ht]
\centering
\subfigure[]{\includegraphics[width=0.49\textwidth]{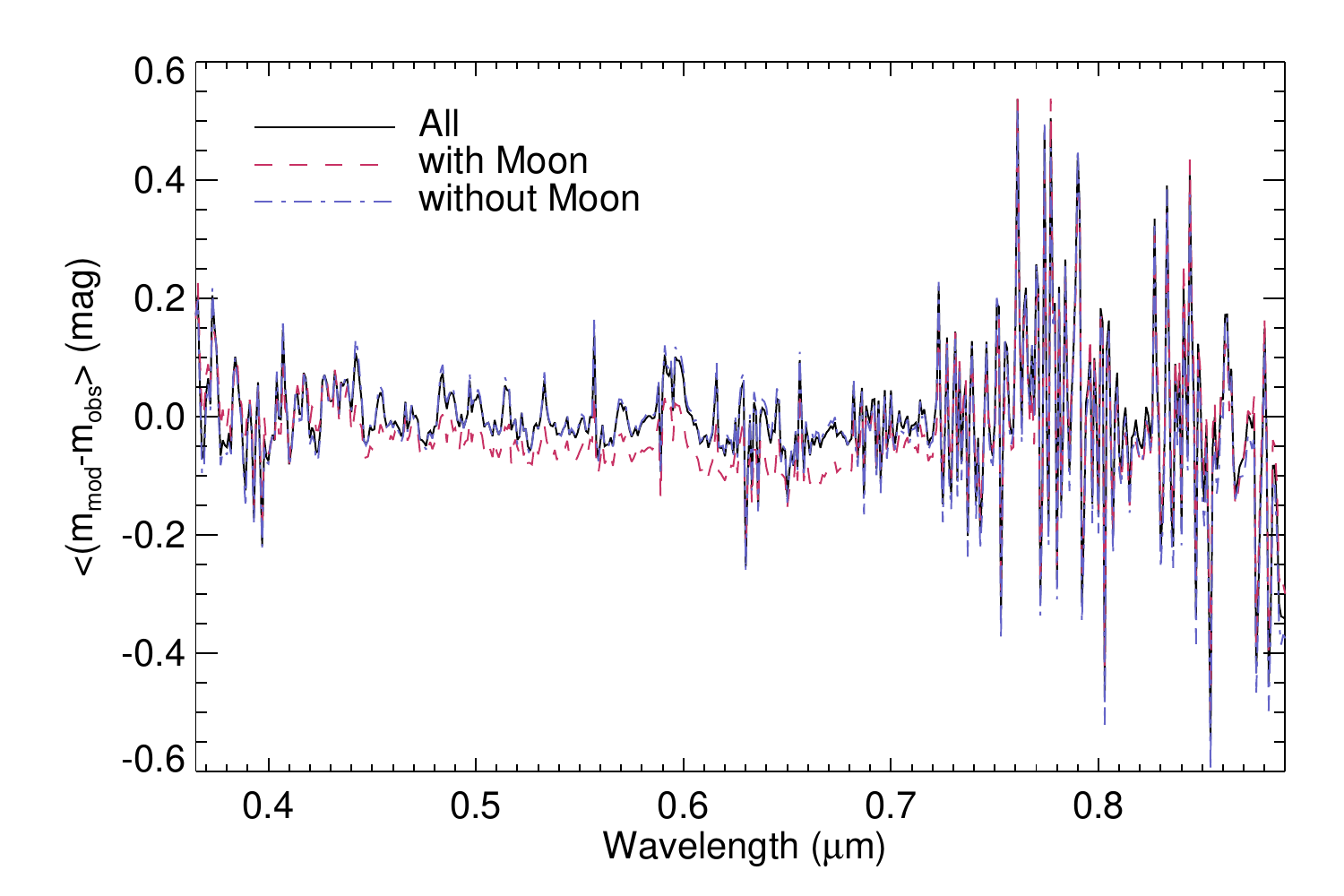}}\hfill
\subfigure[]{\includegraphics[width=0.49\textwidth]{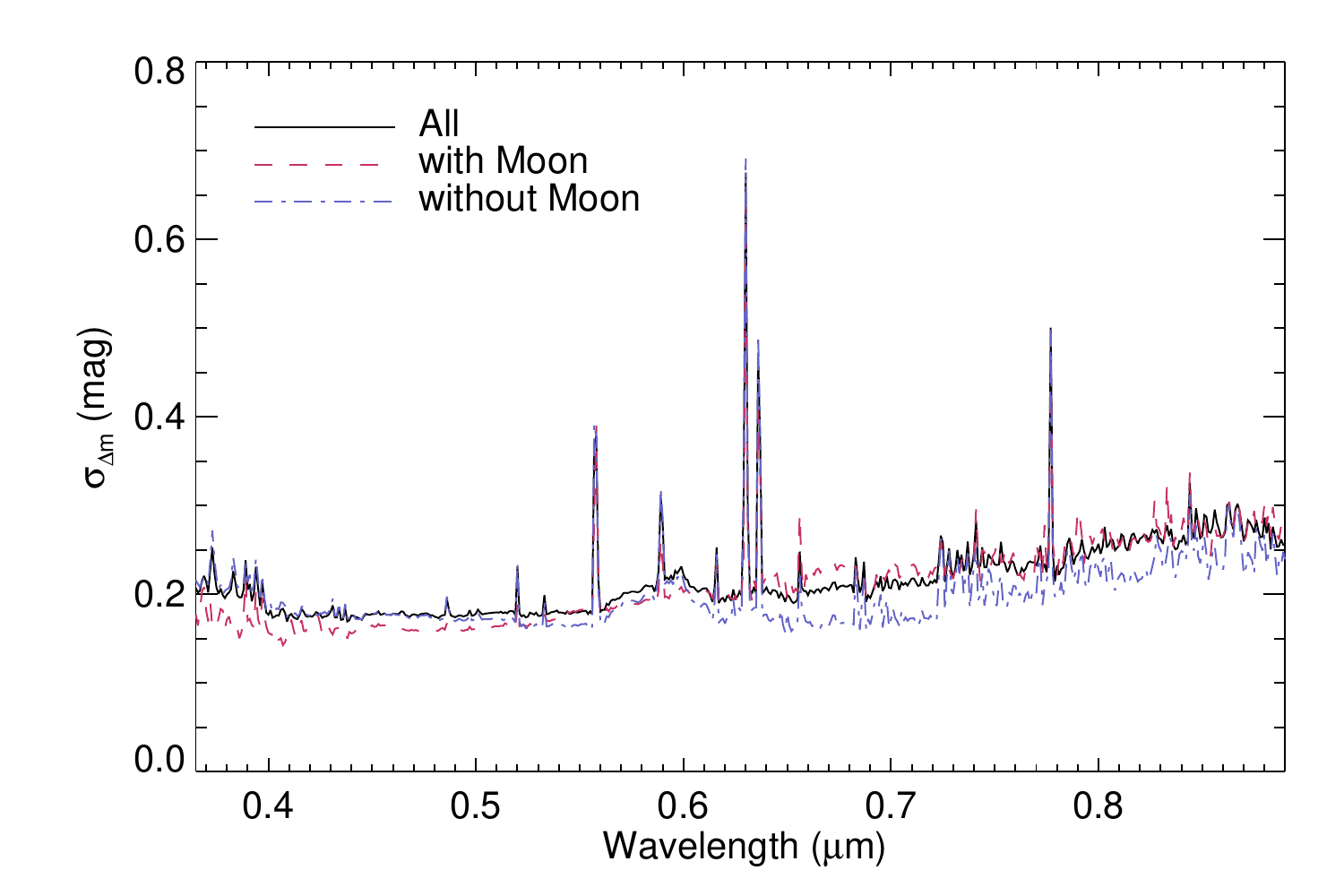}}\\
\caption{Mean {\bf (a)} and $\sigma$ {\bf (b)} of the difference between the observed and modeled spectra for the full sky background of all the spectra, those with moonlight, and only those without moonlight.  See Section 3 for more details.}
\label{plot_spec}
\end{figure}

To better understand the errors of the moon model, we made a mean and $\sigma$ spectrum of the difference between the FORS1 observations and the full sky model for all the spectra, those with moonlight, and those without moonlight (Fig. \ref{plot_spec}).  The mean spectrum is centered around zero, with increased fluctuations at the redder wavelengths due to airglow emission.  The $\sigma$ spectrum without moonlight is slightly lower, but all three groups tend to lie around 0.2 mag, and increase slightly at the red end.  Loosely comparing the spectrum with moonlight to those without, gives a rough estimate of the quality of our advanced moon model.

\begin{figure}[!ht]
\centering
\includegraphics[width=0.49\textwidth]{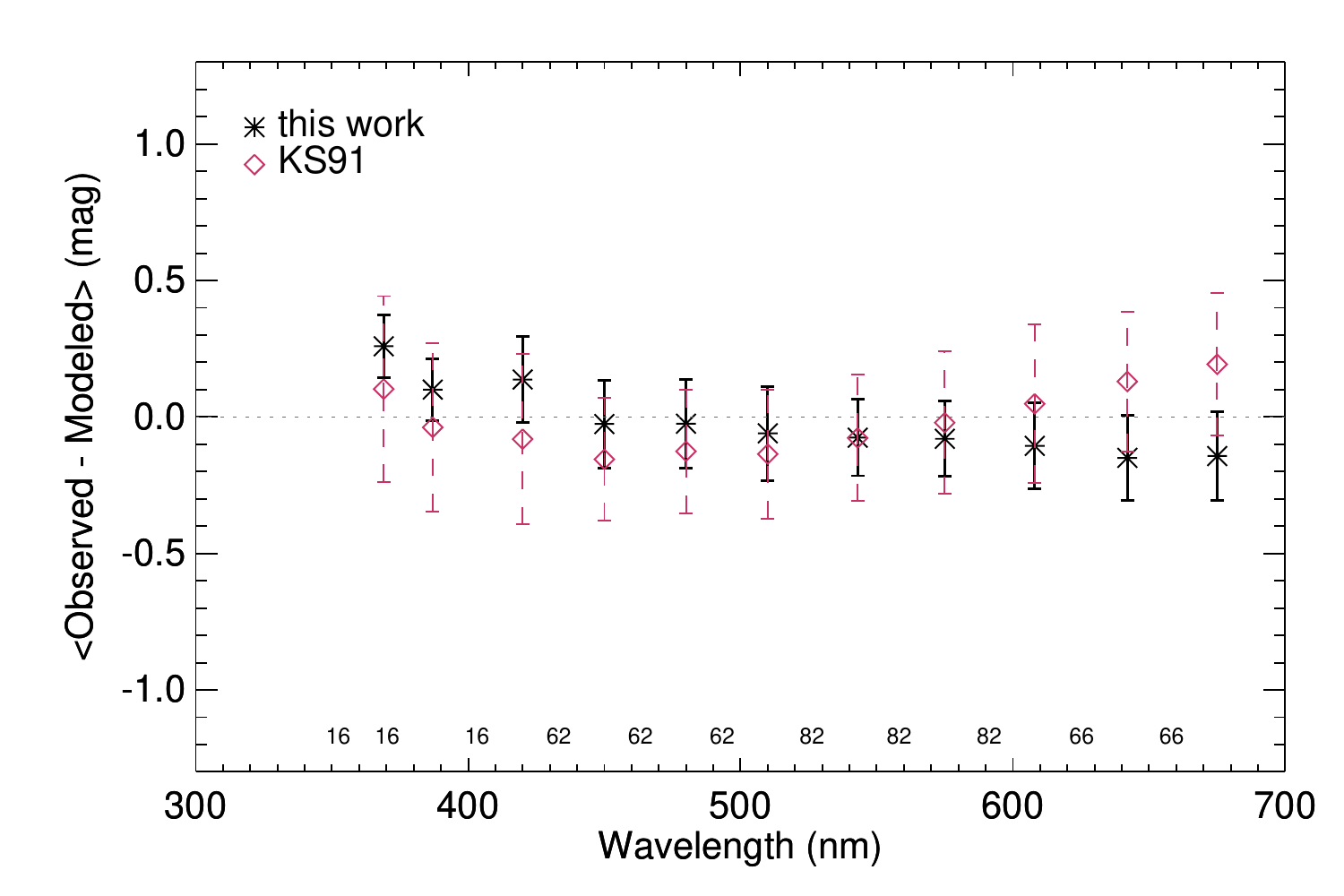}
\caption{Means and uncertainties for the scattered moonlight versus observed data at several 4 nm wide continuum bands.  The y-axis is the average of the observed minus modeled fluxes for data with good weather conditions and a significant amount of moonlight.  Over-plotted is the same analysis with the previous model from \citet{2012A&A...543A..92N} based on \citet{1991PASP..103.1033K}, labeled as KS91. The numbers below each point are the number of spectra considered.  See Section 3 for more details.}
\label{plot_mean}
\end{figure}

To better evaluate only the scattered moonlight model, we subtracted the other components of the sky background, using the sky background model, from the observations.  This, in principle, should provide only an observed scattered moonlight spectrum.  We then compare this observed moon-only spectrum with the scattered moonlight model.  Such a comparison contains all the errors associated with the full sky background model as well as the errors from the moon model.  For an estimate of the uncertainty in the other sky background components, we looked at the full sky model without moonlight.  We then computed a weighted $\sigma$ to represent the uncertainty from the other sky background components, by comparing the relative flux from the moonlight with full sky background for the Moon observations used in this analysis.  If there is less moonlight, then $\sigma$ from the other components should be larger, versus an observation dominated by moonlight where $\sigma$ should mostly come from the moon model.  We did this comparison using all the data with a significant amount of moonlight ($> 100$ phot s$^{-1}\mathrm{m}^{-2}\mu \mathrm{m}^{-1}$arcsec$^{-2}$) and decent weather conditions.  Also, we ignored one of the observing modes, Grism 600B, which had only 8 data points and the flux calibrations appeared to greatly deviate from the other observing modes.  This left us with a total of 82 spectra.  The results can be seen in Fig. \ref{plot_mean}.  There is a jump in the mean around 500 nm where the majority of the observations changes from one observing mode to another.  We believe this comes from errors in the flux calibration of the \citet{2008A&A...481..575P} data.  For the mean of these spectra to be centered at zero, we needed to multiply the moon model by a factor of 1.2 to correct for the flux calibration.  The error bars shown here are the uncertainties from the moon model squared minus the square of the weighted $\sigma$ from the other components.  At blue wavelengths $\sigma$ is small, and it gets larger towards the red.  The estimated weighted $\sigma$ of the other sky background components are inflated in the UV and red wavelengths.  The airglow/residual continuum model probably overestimates the flux, and so the residuals, especially in the red, are large.  This causes the Moon contribution to be underestimated and $\sigma$ to be inflated, so the last three continuum nodes are not shown for this reason.  In Fig. \ref{plot_spec} (a), this effect due to the airglow model can be seen from the downturn in the mean difference spectra.  Also plotted in Fig. \ref{plot_mean} with the same analysis is the previous moon model discussed in \citet{2012A&A...543A..92N}, which is an extension of \citet{1991PASP..103.1033K} moon model and scaled to \citet{2008A&A...481..575P} data, labeled as KS91.  The mean and $\sigma$ for this previous model are worse than our new advanced moon model.  KS91 has a larger $\sigma$ at blue wavelengths and the mean is more off centered at redder wavelengths compared with this work.  The mean and $\sigma$ for our new advanced moon model and the previous one, before being corrected for the errors from the other sky background components are shown in Table \ref{cont}, along with the weighted $\sigma$ from the rest of the sky background.

\begin{figure}[!ht]
\centering
\subfigure[]{\includegraphics[width=0.49\textwidth]{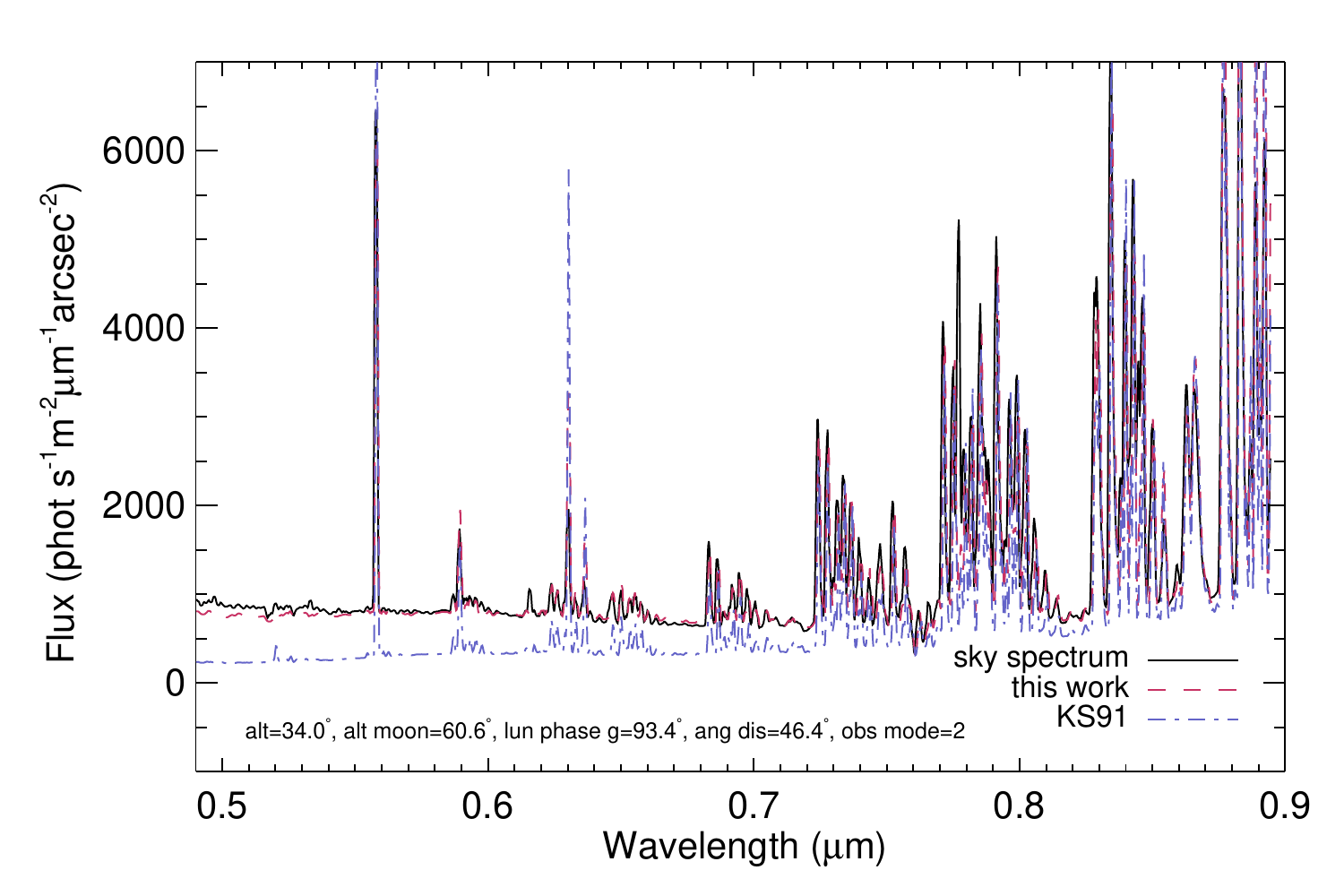}}\hfill
\subfigure[]{\includegraphics[width=0.49\textwidth]{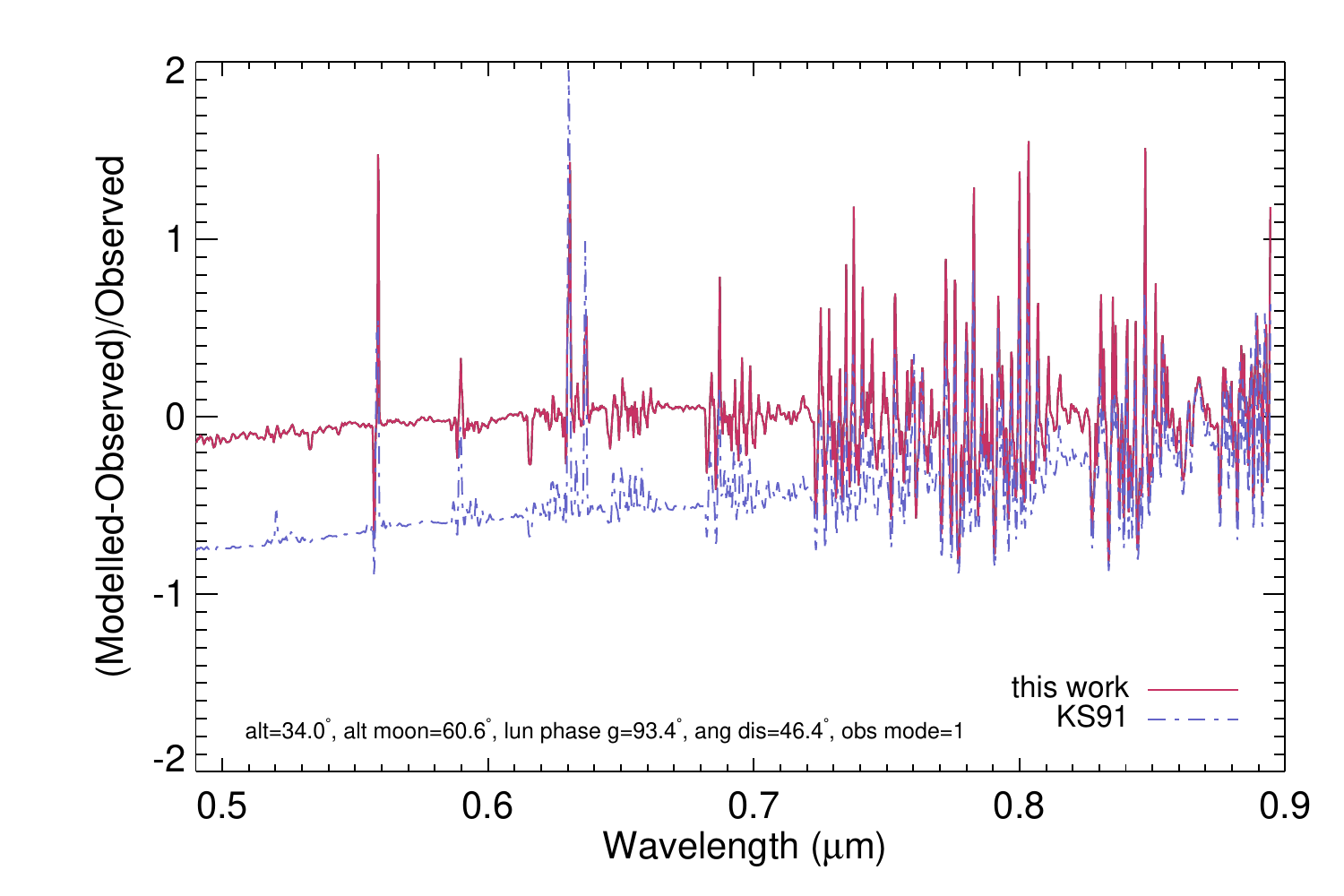}}\\
\caption{Comparison of the full sky background model with the previous moon model based on \citet{1991PASP..103.1033K}, labeled as KS91, and then one presented in this paper.  {\bf (a):}  The observed spectrum along with the full sky background model with the old (KS91) approach and this work. {\bf (b):} This shows the relative residuals for the previous model and the work presented in this paper.  See Section 3 for more details.}
\label{plot_dr05}
\end{figure}

To further show the improvement of our new advanced scattered moonlight model, we have plotted the full sky background model using the extrapolated \citet{1991PASP..103.1033K} moon model (labeled again as KS91) and our new moon model with an observed spectrum in Fig. \ref{plot_dr05} (a).  It can be clearly seen that the new moon model fits the observed spectrum much better than the previous model.  The relative residuals are plotted on panel (b).

\section{Conclusion}

We have developed an advanced scattered moonlight model.  It is based on physical processes and data, and has fewer empirical parametrizations than previous models.  It is spectroscopic and traces spectral trends seen in observations.  Currently, it has been evaluated for the optical range from 0.36 to 0.89 micron.  The uncertainty in the moon model is around 0.15 mag, and the full sky background model has $\sigma \sim 0.2$ mag.  There are several advantages of our new moon model:

$\bullet$ More physical than previous models \citep{1991PASP..103.1033K,2012A&A...543A..92N}

$\quad\diamond$ Uses a fit for the lunar albedo based on more than 100,000 observations of the Moon \citep{2005AJ....129.2887K}

$\quad\diamond$ Uses fully 3-D single scattering calculations

$\quad\diamond$ Calculates fully 3-D double scattering calculations, including ground reflection, then simplifies for faster computing

$\quad\diamond$ Approximates multiple scattering by comparing double and single scattering

$\quad\diamond$ Calculates the absorption and single scattering simultaneously

$\quad\diamond$ Decomposes the Mie scattering \AA ngstr\"{o}m Law into typical aerosol distributions for Cerro Paranal

$\quad\diamond$ Calculates the Mie phase function based on the aerosols found by decomposing the \AA ngstr\"{o}m Law 

$\bullet$ Provides a scattered moonlight spectrum, not just photometric magnitudes \citep{1987Noao....10...16W,1991PASP..103.1033K}

$\bullet$ The uncertainties across the entire optical range are much lower than previous \citep{1991PASP..103.1033K,2012A&A...543A..92N}  

\begin{acknowledgements}
This study is carried out in the framework of the Austrian ESO In-kind project funded by BM:wf under contracts BMWF-10.490/0009-II/10/2009 and BMWF-10.490/0008-II/3/2011.
\end{acknowledgements}

\begin{table}
\caption{Narrow band filter statistics for our new moon model (new), \citet{2012A&A...543A..92N} moon model (old), and the weighted $\sigma$ of the other sky background components (SBC)\label{cont}}
\centering
\begin{tabular}{ccccccc}
\hline\hline
\noalign{\smallskip}
Filter & \# & $<\mathrm{new}>$ & $\sigma_\mathrm{new}$ & $<\mathrm{old}>$ & $\sigma_\mathrm{old}$  & $\sigma_\mathrm{SBC}$ \\
(nm) & & & & & &\\
\noalign{\smallskip}
\hline
\noalign{\smallskip}
369 & 16 & 0.259 & 0.222 & 0.102 & 0.390 & 0.190\\
387 & 16 & 0.100 & 0.213 & -0.038 & 0.357 & 0.180\\
420 & 16 & 0.137 & 0.191 & -0.081 & 0.331 & 0.107\\
450 & 62 & -0.026 & 0.185 & -0.155 & 0.242 & 0.091\\
480 & 62 & -0.025 & 0.192 & -0.126 & 0.248 & 0.101\\
510 & 62 & -0.060 & 0.208 & -0.136 & 0.264 & 0.117\\
543 & 82 & -0.076 & 0.220 & -0.077 & 0.287 & 0.170\\
575 & 82 & -0.080 & 0.299 & -0.021 & 0.372 & 0.265\\
608 & 82 & -0.106 & 0.330 & 0.048 & 0.410 & 0.290\\
642 & 66 & -0.150 & 0.314 & 0.130 & 0.374 & 0.272\\
675 & 66 & -0.143 & 0.326 & 0.193 & 0.384 & 0.283\\
720 & 66 & -0.084 & 0.364 & 0.269 & 0.413 & 0.433\\
820 & 46 & -0.152 & 0.676 & 0.298 & 0.721 & 0.791\\
872 & 45 & 0.016 & 0.761 & 0.483 & 0.818 & 1.029\\
\noalign{\smallskip}
\hline
\end{tabular}
\tablefoot{The listed $\sigma_\mathrm{new}$ and $\sigma_\mathrm{old}$ are before subtracting the uncertainties from the other sky background components.}
\end{table}

\bibliographystyle{aa}
\bibliography{moon}

\begin{thebibliography}{38}
\expandafter\ifx\csname natexlab\endcsname\relax\def\natexlab#1{#1}\fi

\bibitem[{{\AA ngstr\"{o}m}(1929)}]{1929GA.....11..156A}
{\AA ngstr\"{o}m}, A. 1929, Geografiska Annaler, 11, 156

\bibitem[{{Bernstein} {et~al.}(2002){Bernstein}, {Freedman}, \&
  {Madore}}]{2002ApJ...571...56B}
{Bernstein}, R.~A., {Freedman}, W.~L., \& {Madore}, B.~F. 2002, \apj, 571, 56

\bibitem[{{Bohren} \& {Huffman}(1983)}]{1983asls.book.....B}
{Bohren}, C.~F. \& {Huffman}, D.~R. 1983, {Absorption and scattering of light
  by small particles}

\bibitem[{{Buton} {et~al.}(2013){Buton}, {Copin}, {Aldering}, {Antilogus},
  {Aragon}, {Bailey}, {Baltay}, {Bongard}, {Canto}, {Cellier-Holzem},
  {Childress}, {Chotard}, {Fakhouri}, {Gangler}, {Guy}, {Hsiao}, {Kerschhaggl},
  {Kowalski}, {Loken}, {Nugent}, {Paech}, {Pain}, {P{\'e}contal}, {Pereira},
  {Perlmutter}, {Rabinowitz}, {Rigault}, {Runge}, {Scalzo}, {Smadja}, {Tao},
  {Thomas}, {Weaver}, {Wu}, \& {Nearby SuperNova
  Factory}}]{2013A&A...549A...8B}
{Buton}, C., {Copin}, Y., {Aldering}, G., {et~al.} 2013, \aap, 549, A8

\bibitem[{{Clough} {et~al.}(2005){Clough}, {Shephard}, {Mlawer}, {Delamere},
  {Iacono}, {Cady-Pereira}, {Boukabara}, \& {Brown}}]{2005JQSRT..91..233C}
{Clough}, S.~A., {Shephard}, M.~W., {Mlawer}, E.~J., {et~al.} 2005, \jqsrt, 91,
  233

\bibitem[{{Colina} {et~al.}(1996){Colina}, {Bohlin}, \&
  {Castelli}}]{1996AJ....112..307C}
{Colina}, L., {Bohlin}, R.~C., \& {Castelli}, F. 1996, \aj, 112, 307

\bibitem[{{Davies} {et~al.}(2013){Davies}, {Bennie}, {Inger}, \&
  {Gaston}}]{2013NatSR...3E1722D}
{Davies}, T.~W., {Bennie}, J., {Inger}, R., \& {Gaston}, K.~J. 2013, Scientific
  Reports, 3

\bibitem[{{Dollfus} \& {Bowell}(1971)}]{1971A&A....10...29D}
{Dollfus}, A. \& {Bowell}, E. 1971, \aap, 10, 29

\bibitem[{{Elterman}(1966)}]{1966ApOpt...5.1769E}
{Elterman}, L. 1966, \ao, 5, 1769

\bibitem[{{G{\'a}l} {et~al.}(2001){G{\'a}l}, {Horv{\'a}th}, {Barta}, \&
  {Wehner}}]{2001JGR...10622647G}
{G{\'a}l}, J., {Horv{\'a}th}, G., {Barta}, A., \& {Wehner}, R. 2001, \jgr, 106,
  22647

\bibitem[{{Grainger} {et~al.}(2004){Grainger}, {Lucas}, {Thomas}, \&
  {Ewen}}]{2004ApOpt..43.5386G}
{Grainger}, R.~G., {Lucas}, J., {Thomas}, G.~E., \& {Ewen}, G.~B.~L. 2004, \ao,
  43, 5386

\bibitem[{{Henyey} \& {Greenstein}(1941)}]{1941ApJ....93...70H}
{Henyey}, L.~G. \& {Greenstein}, J.~L. 1941, \apj, 93, 70

\bibitem[{{Horv{\'a}th} {et~al.}(1998){Horv{\'a}th}, {G{\'a}l}, {Pomozi}, \&
  {Wehner}}]{1998NW.....85..333H}
{Horv{\'a}th}, G., {G{\'a}l}, J., {Pomozi}, I., \& {Wehner}, R. 1998,
  Naturwissenschaften, 85, 333

\bibitem[{{Horvath} {et~al.}(2006){Horvath}, {Kasahara}, {Tohno}, \&
  {Kocifaj}}]{2006AS..37..1287H}
{Horvath}, H., {Kasahara}, M., {Tohno}, S., \& {Kocifaj}, M. 2006, Aerosol
  Science, 37, 1287

\bibitem[{{Kieffer} \& {Stone}(2005)}]{2005AJ....129.2887K}
{Kieffer}, H.~H. \& {Stone}, T.~C. 2005, \aj, 129, 2887

\bibitem[{{Kleipool} {et~al.}(2008){Kleipool}, {Dobber}, {de Haan}, \&
  {Levelt}}]{2008JGRD..11318308K}
{Kleipool}, Q.~L., {Dobber}, M.~R., {de Haan}, J.~F., \& {Levelt}, P.~F. 2008,
  Journal of Geophysical Research (Atmospheres), 113, 18308

\bibitem[{{Knoetig} {et~al.}(2013){Knoetig}, {Biland}, {Bretz}, {Bu{\ss}},
  {Dorner}, {Einecke}, {Eisenacher}, {Hildebrand}, {Kr{\"a}henb{\"u}hl},
  {Lustermann}, {Mannheim}, {Meier}, {Neise}, {Overkemping}, {Paravac},
  {Pauss}, {Rhode}, {Ribordy}, {Steinbring}, {Temme}, {Thaele}, {Vogler},
  {Walter}, {Weitzel}, \& {Z{\"a}nglein}}]{2013arXiv1307.6116K}
{Knoetig}, M.~L., {Biland}, A., {Bretz}, T., {et~al.} 2013, ArXiv e-prints

\bibitem[{{Krisciunas} \& {Schaefer}(1991)}]{1991PASP..103.1033K}
{Krisciunas}, K. \& {Schaefer}, B.~E. 1991, \pasp, 103, 1033

\bibitem[{{Lane} \& {Irvine}(1973)}]{1973AJ.....78..267L}
{Lane}, A.~P. \& {Irvine}, W.~M. 1973, \aj, 78, 267

\bibitem[{{Leinert} {et~al.}(1998){Leinert}, {Bowyer}, {Haikala}, {Hanner},
  {Hauser}, {Levasseur-Regourd}, {Mann}, {Mattila}, {Reach}, {Schlosser},
  {Staude}, {Toller}, {Weiland}, {Weinberg}, \& {Witt}}]{1998A&AS..127....1L}
{Leinert}, C., {Bowyer}, S., {Haikala}, L.~K., {et~al.} 1998, \aaps, 127, 1

\bibitem[{{Levelt} {et~al.}(2006){Levelt}, {van den Oord}, {Dobber}, {Malkki},
  {Visser}, {de Vries}, {Stammes}, {Lundell}, \& {Saari}}]{2006ITGRS..44.1093L}
{Levelt}, P.~F., {van den Oord}, G.~H.~J., {Dobber}, M.~R., {et~al.} 2006, IEEE
  Transactions on Geoscience and Remote Sensing, 44, 1093

\bibitem[{{Liou}(2002)}]{2002aiar.book.....L}
{Liou}, K.~N. 2002, {An introduction to atmospheric radiation, 2nd ed}

\bibitem[{{Mayer} \& {Kylling}(2005)}]{2005ACP.....5.1855M}
{Mayer}, B. \& {Kylling}, A. 2005, Atmospheric Chemistry \& Physics, 5, 1855

\bibitem[{{Noll} {et~al.}(2012){Noll}, {Kausch}, {Barden}, {Jones}, {Szyszka},
  {Kimeswenger}, \& {Vinther}}]{2012A&A...543A..92N}
{Noll}, S., {Kausch}, W., {Barden}, M., {et~al.} 2012, \aap, 543, A92

\bibitem[{{Patat}(2008)}]{2008A&A...481..575P}
{Patat}, F. 2008, \aap, 481, 575

\bibitem[{{Patat} {et~al.}(2011){Patat}, {Moehler}, {O'Brien}, {Pompei},
  {Bensby}, {Carraro}, {de Ugarte Postigo}, {Fox}, {Gavignaud}, {James},
  {Korhonen}, {Ledoux}, {Randall}, {Sana}, {Smoker}, {Stefl}, \&
  {Szeifert}}]{2011A&A...527A..91P}
{Patat}, F., {Moehler}, S., {O'Brien}, K., {et~al.} 2011, \aap, 527, A91

\bibitem[{{Rothman} {et~al.}(2009){Rothman}, {Gordon}, {Barbe}, {Benner},
  {Bernath}, {Birk}, {Boudon}, {Brown}, {Campargue}, {Champion}, {Chance},
  {Coudert}, {Dana}, {Devi}, {Fally}, {Flaud}, {Gamache}, {Goldman},
  {Jacquemart}, {Kleiner}, {Lacome}, {Lafferty}, {Mandin}, {Massie},
  {Mikhailenko}, {Miller}, {Moazzen-Ahmadi}, {Naumenko}, {Nikitin}, {Orphal},
  {Perevalov}, {Perrin}, {Predoi-Cross}, {Rinsland}, {Rotger}, {{\v S}ime{\v
  c}kov{\'a}}, {Smith}, {Sung}, {Tashkun}, {Tennyson}, {Toth}, {Vandaele}, \&
  {Vander Auwera}}]{2009JQSRT.110..533R}
{Rothman}, L.~S., {Gordon}, I.~E., {Barbe}, A., {et~al.} 2009, \jqsrt, 110, 533

\bibitem[{{Seifahrt} {et~al.}(2010){Seifahrt}, {K{\"a}ufl}, {Z{\"a}ngl},
  {Bean}, {Richter}, \& {Siebenmorgen}}]{2010A&A...524A..11S}
{Seifahrt}, A., {K{\"a}ufl}, H.~U., {Z{\"a}ngl}, G., {et~al.} 2010, \aap, 524,
  A11

\bibitem[{{Staude}(1975)}]{1975A&A....39..325S}
{Staude}, H.~J. 1975, \aap, 39, 325

\bibitem[{{Sutter} {et~al.}(2007){Sutter}, {Dalton}, {Ewing}, {Amundson}, \&
  {McKay}}]{2007JGRG..112.4S10S}
{Sutter}, B., {Dalton}, J.~B., {Ewing}, S.~A., {Amundson}, R., \& {McKay},
  C.~P. 2007, Journal of Geophysical Research (Biogeosciences), 112, 4

\bibitem[{{Trinh} {et~al.}(2013){Trinh}, {Ellis}, {Bland-Hawthorn}, {Horton},
  {Lawrence}, \& {Leon-Saval}}]{2013MNRAS.432.3262T}
{Trinh}, C.~Q., {Ellis}, S.~C., {Bland-Hawthorn}, J., {et~al.} 2013, \mnras,
  432, 3262

\bibitem[{{Velikodsky} {et~al.}(2011){Velikodsky}, {Opanasenko}, {Akimov},
  {Korokhin}, {Shkuratov}, {Kaydash}, {Videen}, {Ehgamberdiev}, \&
  {Berdalieva}}]{2011Icar..214...30V}
{Velikodsky}, Y.~I., {Opanasenko}, N.~V., {Akimov}, L.~A., {et~al.} 2011,
  Icarus, 214, 30

\bibitem[{{Vernet} {et~al.}(2011){Vernet}, {Dekker}, {D'Odorico}, {Kaper},
  {Kjaergaard}, {Hammer}, {Randich}, {Zerbi}, {Groot}, {Hjorth}, {Guinouard},
  {Navarro}, {Adolfse}, {Albers}, {Amans}, {Andersen}, {Andersen}, {Binetruy},
  {Bristow}, {Castillo}, {Chemla}, {Christensen}, {Conconi}, {Conzelmann},
  {Dam}, {de Caprio}, {de Ugarte Postigo}, {Delabre}, {di Marcantonio},
  {Downing}, {Elswijk}, {Finger}, {Fischer}, {Flores}, {Fran{\c c}ois},
  {Goldoni}, {Guglielmi}, {Haigron}, {Hanenburg}, {Hendriks}, {Horrobin},
  {Horville}, {Jessen}, {Kerber}, {Kern}, {Kiekebusch}, {Kleszcz}, {Klougart},
  {Kragt}, {Larsen}, {Lizon}, {Lucuix}, {Mainieri}, {Manuputy}, {Martayan},
  {Mason}, {Mazzoleni}, {Michaelsen}, {Modigliani}, {Moehler}, {M{\o}ller},
  {Norup S{\o}rensen}, {N{\o}rregaard}, {P{\'e}roux}, {Patat}, {Pena}, {Pragt},
  {Reinero}, {Rigal}, {Riva}, {Roelfsema}, {Royer}, {Sacco}, {Santin},
  {Schoenmaker}, {Spano}, {Sweers}, {Ter Horst}, {Tintori}, {Tromp}, {van
  Dael}, {van der Vliet}, {Venema}, {Vidali}, {Vinther}, {Vola}, {Winters},
  {Wistisen}, {Wulterkens}, \& {Zacchei}}]{2011A&A...536A.105V}
{Vernet}, J., {Dekker}, H., {D'Odorico}, S., {et~al.} 2011, \aap, 536, A105

\bibitem[{{Walker}(1987)}]{1987Noao....10...16W}
{Walker}, A. 1987, NOAO Newsletter, 10, 16

\bibitem[{{Wallace} {et~al.}(2011){Wallace}, {Hinkle}, {Livingston}, \&
  {Davis}}]{2011ApJS..195....6W}
{Wallace}, L., {Hinkle}, K.~H., {Livingston}, W.~C., \& {Davis}, S.~P. 2011,
  \apjs, 195, 6

\bibitem[{{Warneck} \& {Williams}(2012)}]{2012acc.book.....W}
{Warneck}, P. \& {Williams}, J. 2012, {The Atmospheric Chemist's Companion.},
  ed. {Springer}

\bibitem[{{Whitaker}(1969)}]{1969NASSP.201...38W}
{Whitaker}, E.~A. 1969, NASA Special Publication, 201, 38

\bibitem[{{Wolstencroft} \& {van Breda}(1967)}]{1967ApJ...147..255W}
{Wolstencroft}, R.~D. \& {van Breda}, I.~G. 1967, \apj, 147, 255

\end{thebibliography}

\end{document}